\documentclass[onecolumn,showpacs]{revtex4}
\usepackage{graphicx}
\usepackage{epsfig}

\newcommand{\beq}{\begin{equation}}
\newcommand{\eeq}{\end{equation}}

\begin{document}

\title{From Feshbach-Resonance Managed Bose-Einstein  
Condensates to Anisotropic Universes: \\
Applications of the Ermakov-Pinney equation
with Time-Dependent Nonlinearity }
\author{G. Herring$^1$, P.G. Kevrekidis$^1$, F. Williams$^1$, 
T. Christodoulakis$^2$ and D.J. Frantzeskakis$^2$}
\affiliation{ 
$^{1}$ Department of Mathematics and Statistics,
University of Massachusetts, Amherst MA 01003-4515, USA \\
$^{2}$ 
Department of Physics, University of Athens,
Panepistimiopolis, Zografos, Athens 15784, Greece}

\begin{abstract}
In this work 
we revisit the topic of two-dimensional
Bose-Einstein condensates under the influence of time-dependent 
magnetic confinement and time-dependent scattering length. 
A moment approach reduces the examination of moments of the
wavefunction (in particular, of its width) to an Ermakov-Pinney (EP)
ordinary differential equation (ODE). We use the well-known structure
of the solutions of this nonlinear ODE to ``engineer'' trapping and
interatomic interaction conditions that lead to condensates dispersing,
breathing or even collapsing. The advantage of the approach is that
it is fully tractable analytically, in excellent agreement with
our numerical observations. As an aside, we also discuss how similar
time-dependent EP equations may arise in the description of anisotropic
scalar field cosmologies.
\end{abstract}

\maketitle


\section{Introduction}

The recent progress in experimental and theoretical studies of Bose-Einstein
condensates (BECs) of dilute atomic gases \cite{review} has been tremendous after their experimental
realization \cite{prop}. This has also led to an explosion of interest in the theme of 
nonlinear matter-waves 
such as dark \cite{dark}, bright \cite{bright} and gap \cite{gap} solitons.
Two-dimensional (2D) nonlinear excitations of BECs, such as 
vortices \cite{vortex} and vortex lattices \cite{vl}, were also 
realized experimentally, while 
a multitude of 
other coherent nonlinear structures were also 
theoretically predicted; these include, e.g., Faraday waves \cite{stal}, 
ring dark solitons and vortex necklaces \cite{theo}, 
stable solitons and localized vortices in attractive BECs trapped in  
periodic potentials \cite{Salerno}, 
matter-wave gap vortices 
\cite{lena}, 2D bright solitons in dipolar BECs \cite{santos}, and so on.

From the theoretical standpoint, the dynamics of such higher-dimensional structures 
is, generally, difficult to be treated analytically,
especially so in time-dependent settings. This, in turn, raises important 
questions concerning their ``controllability'', which would be of particular 
relevance regarding potential applications. 
Ideally, such a controllability 
would allow 
``manipulation'' of the condensates at will, e.g., 
sustaining condensates which may expand towards 
complete dispersion, contract towards a complete collapse, or
perform stable breathing oscillations. Moreover, such processes could involve 
a targeted growth of a condensate up to 
a certain width or shrinkage down a desired size. Our purpose
in the present work 
is to illustrate how one can achieve this goal, by taking 
advantage of one of the few analytically tractable
tools in higher-dimensional condensates, namely of the analysis
of their moments \cite{vic2,vic3,krb1} (see also references 
therein). In particular, we focus on the case of quasi 
two-dimensional (2D), so-called, pancake-shaped condensates \cite{pancake},
under the effect of 
time-varying 
harmonic trapping \cite{vic3,krb1} 
and also time-dependent s-wave scattering length (which controls the inter-particle interactions). 
Note that the controllability over the harmonic  
trapping is straightforwardly realizable under AC-variation of the atom trapping frequencies, while 
the controllabity over the interatomic interactions   
can be realized by using the so-called Feshbach resonance \cite{inouye}, 
connecting the s-wave scattering length 
to external magnetic fields.
This latter 
technique (which is usually called Feshbach resonance management (FRM)) has been proposed theoretically
as a means of avoiding collapse \cite{bor1,bor2,vic1}, but also as a way 
of producing robust coherent nonlinear matter-waves 
\cite{our1,our2,our3,kon}.

For the quasi-2D 
condensates discussed above,
upon presenting the moment analysis, we obtain a 
reduced dynamical description effectively involving only
a variable associated with the width of the condensate 
wavefunction. This ordinary differential equation (ODE) 
can be reduced to one of the Ermakov-Pinney (EP) type 
\cite{hawkins,hawk,espin,floyd}, whose solutions can be 
obtained {\it analytically}, provided that the solutions 
of the underlying linear Schr{\"o}dinger equation can be 
obtained. We use this feature and the freedom in selecting 
the time-dependent trapping and interactions of the condensates 
to illustrate that one can construct analytical solutions 
to this ODE 
that completely characterize the temporal evolution of the width of the 
wavefunction. In so doing, we fully prescribe the 
dynamical evolution of the condensate. We show three 
prototypical examples in applying this idea: one in 
which the width grows in time (leading to condensate 
expansion), one in which it decreases (leading to 
focusing), and one in which it periodically ``breathes'' 
between a minimum and a maximum value. In all three 
cases, we test the analytical prediction against the 
full numerical simulation of the mean-field partial 
differential equation (PDE) model fully describing the 
condensate. We find excellent agreement between the two, 
showcasing the accuracy of our theoretical approach. 

The analysis of the ensuing EP equation in this setting
is of interest in its own right. This is because, as we will
see, the resulting EP equation has a time-dependent nonlinearity
in the right hand side (contrary to what is the case for
the ``standard'' EP framework; see e.g. \cite{hawkins,hawk,espin,floyd}
and references therein). In this context, and as an aside,
we present a second physically relevant example where
such EP equations with time-dependent nonlinearities
may arise, by studying anisotropic scalar field cosmologies
of a particular anisotropic geometry. This generalizes the
example of \cite{hawkins} parallelizing BECs without FRM and scalar field
cosmologies in the isotropic case.

Our presentation will be structured as follows: In 
section III, we provide a brief synopsis of the main 
features of the EP equation. 
In section IV, we proceed to an overview 
of the moment analysis, following the earlier works 
\cite{vic2,vic3,krb1}. In section IV, we present 
the analytical solutions of the EP equation that we develop 
for each of the above-mentioned three cases. Section V 
tests these analytical results against full simulations of 
the PDE describing the condensate. 
Section VI illustrates the second example of time-dependent
EP equations in the description of anisotropic scalar field
cosmologies.
Finally, in section VII,
we summarize our findings and present our conclusions.

\section{The Ermakov-Pinney Equation}

The Ermakov-Pinney (EP) equation is a remarkable nonlinear ODE of the form:
\begin{eqnarray}
Y''+  Q(\tau) Y= \frac{\kappa}{Y^3}.
\label{oeq0}
\end{eqnarray}
The particularly attractive feature of this nonlinear ODE is that its 
general solution can be obtained, provided that one is able
to solve the time-independent linear Schr{\"o}dinger (LS) 
equation $Y'' + Q(\tau) Y=0$.
For details on the properties of the EP equation, 
the interested reader is referred to 
\cite{hawk,espin} and references therein. Here we just mention its basic 
superposition principle property. Namely, if the linearly independent 
solutions of the LS equation are $Y_1(\tau)$ and $Y_2(\tau)$, then the 
most general possible solution of the EP equation is given by
\begin{eqnarray}
Y(\tau)=\left( A Y_1^2 + B Y_2^2 + 2 C Y_1 Y_2 \right)^{1/2}
\label{oeq8a}
\end{eqnarray}
where $A, B$ and $C$ are constants connected through
\begin{eqnarray}
A B - C^2= \frac{\kappa}{W^2}
\label{oeq8b}
\end{eqnarray}
%
where $W= Y_1 Y_2' - Y_2 Y_1'$ is the Wronskian of $Y_1(\tau)$ and $Y_2(\tau)$.

\section{Moment Analysis For BECs}

One of the interesting variants of the ``regular'' EP equation
of the form (\ref{oeq0}) arises in the study of BECs, 
albeit in a somewhat modified form (see below). 

The relevant mean-field model for studying atomic Bose-Einstein condensates at zero temperature 
consists of the so-called Gross-Pitaevskii equation \cite{review} of the following dimensionless form:
\begin{eqnarray}
i \partial_{t} u=-\frac{1}{2} \Delta u+ \left (\lambda(t) r^2+ \nu(t) |u|^2\right) u,
\label{eq1}
\end{eqnarray}
where $u$ represents the wavefunction of the condensate, $V(r)=\lambda(t) r^2$ 
denotes the harmonic 
trap confining the bosons [with a time-dependent frequency determined by $\lambda(t)$]
and $\nu(t)$ is the 
coefficient of the nonlinear term proportional to the s-wave scattering length, 
characterizing the interaction between the particles. We will take advantage 
of the Feshbach resonance \cite{inouye} to 
consider that the latter is time-dependent, as well. 

One of the popular approaches to studying Eq. (\ref{eq1}) is through
the use of moment methods \cite{vic2,vic3,krb1}. 
The latter allow us to write ODEs for the moments of the
mean-field wavefunction $u$ 
as follows. We define 
\begin{eqnarray}
I_{2,a}^{(d)}&=&\int_0^\infty r^a |u|^2r^{d-1}dr,
\label{eq3}
\\
I_{3,a}^{(d)}&=&i \int_0^\infty r^a (u u^{\star}_r - u^{\star} u_r)
r^{d-1}dr,
\label{eq4}
\\
I_{4,a}^{(d)}&=&\int_0^\infty  r^a \left |\frac{\partial u}{\partial r}\right |^2 r^{d-1}dr,
\label{eq5}
\\
I_{5,a}^{(d)}&=&\int_0^\infty r^a |u|^4 r^{d-1}dr,
\label{eq6}
\end{eqnarray}
where subscripts denote partial differentiation, ``$\star$ denotes complex conjugate and 
$(d)$ indexes the dimension (similarly to \cite{vic2,krb1}). 
The Hamiltonian of Eq. (\ref{eq1})
\begin{eqnarray}
H = \frac{1}{2} \int_0^\infty \left [ | \nabla u|^2+ \nu(t) |u|^4 + 2 \lambda(t) r^2 |u|^2 \right ] r^{d-1}dr,
\label{ham}
\end{eqnarray} 
can then be written as:
\begin{eqnarray}
H=\frac{1}{2} I_{4,0}^{(d)}+ \frac{1}{2} \nu(t) I_{5,0}^{(d)}+\lambda(t)I_{2,2}^{(d)}.
\label{ham1}
\end{eqnarray} 
One can then infer from the dynamics of Eq. (\ref{eq1}) that:
\begin{eqnarray}
\dot{H}=\frac{\nu'(t)}{2} I_{5,0}^{(d)} + {\lambda'(t)}
I_{2,2}^{(d)}
\label{deriv}
\end{eqnarray}
However, as derived in \cite{vic2,krb1},
\begin{equation}
\ddot I^{(d)}_{2,2}= 4 H- 8 \lambda(t)I^{(d)}_{2,2}+2 \nu(t) (d-2)I_{5,0}^{(d)}.
\label{virial}
\end{equation}
We focus on $I_{2,2}^{(d)}$, since this moment 
is associated with the width of the spatial profile of 
the wavefunction. From the latter, we can infer a number of useful pieces
of information concerning certain asymptotic values; in particular, if $I_{2,2}^{(d)} \rightarrow 0$, 
the condensate collapses, while if 
$I_{2,2}^{(d)} \rightarrow  \infty$, the BEC disperses. Furthermore, 
since the $L^2$ norm of the wavefunction is conserved, the estimate
of $I_{2,2}^{(d)}$ on the (square) width of the wavefunction can be
used together with this conservation law to provide information on 
the amplitude $A$ of the wavefunction (i.e., approximately $A \sim 
1/I_{2,2}^{(d)}$).

Combining the two equations (\ref{deriv}) and (\ref{virial}), 
we obtain a single equation for the time-dependence of $I_{2,2}^{(d)}$ as 
\begin{eqnarray}
\dddot{I}_{2,2}^{(d)}=2 \nu'(t) I_{5,0}^{(d)} -4 \lambda'(t) I_{2,2}^{(d)}
- 8 \lambda(t) \dot{I}_{2,2}^{(d)} + \left(d-2\right) \left[ 2 \nu'(t) I_{5,0}^{(d)}
+2 \nu(t) \dot{I}_{5,0}^{(d)} \right]
\label{three}
\end{eqnarray}

Following \cite{vic1}, we consider a quadratic phase for the solution (an assumption most relevant
to dimension $d=2$ as discussed in \cite{vic1}, but which we will also consider more generally), 
and obtain (cf. Eq. (6d) of \cite{vic1}) that 
\begin{eqnarray}
I_{5,0}^{(d)}=\frac{K}{I_{2,2}^{(d)}},
\label{proportionality}
\end{eqnarray}
where the constant $K$ is determined by initial conditions. 
Notice that this is the {\it only} assumption in our 
calculations herein, whose validity will be examined a posteriori
by comparing our analytical results with numerical computations.

Denoting for simplicity $I_{2,2}^{(d)}=y$, the resulting ordinary
differential equation for $y$ can be written as:
\begin{eqnarray}
\frac{d}{dt}\left( y \ddot{y}- \frac{1}{2} \dot{y}^2 - 2 K \nu(t)
+ 4 \lambda(t) y^2 \right)= 2 K (d-2) y \frac{d}{dt}\left(\frac{\nu(t)}{y}\right)
\label{ode}
\end{eqnarray}

Clearly, from the above exposition, the most straightforward case
is the one with $d=2$, on which we will focus next. When $d=2$, the
equation can be directly integrated, yielding
\begin{eqnarray} 
\ddot{y}-\frac{1}{2 y}  \dot{y}^2 +4 \lambda(t) y= \frac{2 K \nu(t) + C}{y},
\label{ode1}
\end{eqnarray}
where $C$ is an integration constant that can be computed from the
initial conditions as:
\begin{eqnarray}
C= 4 I_{2,2}^{(2)}(0) \left( H(0) -  \lambda(0)I^{(2)}_{2,2}(0) \right)
-\frac{1}{2} (I_{3,1}^{(2)}(0))^2
-2 K \nu(0)
\label{ode2}
\end{eqnarray}
 
If we now use the transformation $Y(t)=\sqrt{y(t)}$ \cite{vic3,hawkins},
then we obtain an Ermakov-Pinney (EP) type equation of the form:
\begin{eqnarray}
\ddot{Y}+ 2 \lambda(t) Y= \frac{K \nu(t)+ \frac{C}{2}}{Y^3}
\label{ode3}
\end{eqnarray}
Hence, Eq. (\ref{ode3}) is the equation that describes
the dynamics of two-dimensional Bose-Einstein condensates in the presence
of a time-dependent trap \cite{vic2,vic3,hawkins}, as well as in the
scenario of Feshbach resonance management \cite{bor1,bor2,vic1,our1,our2,our3}.


One can now examine particular cases of time-dependence of 
$\lambda(t)$ and $\nu(t)$, using the analytical tractability
of the ensuing EP equation, in order to obtain completely analytical
solutions for $Y(t)$, and hence for $I_{2,2}^{(2)}$.

\section{Analytical Results}

There are numerous possiblities in the case of a time-dependent 
$\lambda$ and $\nu$, thus, as explained above, we limited our 
investigation to the three cases of an expanding waveform, a
collapsing waveform and an oscillatory waveform. In order to derive 
examples for each of these cases, $\nu(t)$ was chosen to be independent 
of the choice of $\lambda(t)$ and $y(t)$. Solutions to 
Eqn. (\ref{ode3}) were determined by ``reverse engineering'' \cite{floyd}
(based on the desired behavior of the dynamics of the wavefunction), 
and the details of the EP functional form.  Below are the three cases 
examined. Notice that in all three cases, the function 
$f(t)$ used below is given by $f(t)=K \nu(t) + C/2$ i.e., by the
numerator of the right hand side of Eq. (\ref{ode3}).

\begin{enumerate}

\item
The first case explores the possibility of an expanding width wavefunction. In 
order for the wavefunction width to increase, $\lambda(t)$ must be 
decreasing, thus test functions for $\lambda(t)$ and $y(t)$ were chosen to 
reflect this. A similar methodology was also used in the two other cases to 
determine test functions.
\begin{eqnarray}
y(t) &=& B^2\left( {A^2  + t^2 } \right)
\label{Case1a}
\\
\lambda(t) &=& \frac{{f(t) - A^2B^4}}{{2B^4\left( {A^2+t^2} \right)^2 }}
\label{Case1b}
\\
z(t) &=& B^4 A^2 = {\rm const.}
\label{Case1c}
\end{eqnarray}

\item
The second case, a focusing wavefunction, required an additional condition 
($y(0)\neq \infty$) in order to perform the numerical simulations 
described below. Additionally, in this case and the next, an initially
undetermined function $z(t)$ was used in $\lambda(t)$ to simplify the derivation of 
$\lambda(t)$ and $y(t)$. After these were found, the exact form of $z(t)$ 
was calculated by plugging $\lambda(t)$ and $y(t)$ into Eqn. 
(\ref{ode3}) and solving for $z(t)$. This results in
\begin{eqnarray}
y(t) &=& \frac{B^2}{{A^2+t^p}}
\label{Case2a}
\\
\lambda(t) &=& \left[ {f(t) - z(t)} \right] \frac{{B^q }}
{{\left[ 2(A^2  + t^p)  \right]^{q/2} }}
\label{Case2b}
\\
z(t) &=& \frac{{B^5 }} {{(A^2  + t^p )}}\left[ {\frac{{B^4 pt^{p - 2} (pt^p  - 2pA^2  + 2A^2  + 2t^p }} 
{{4(A^2  + t^p )^4 }} + \frac{{f(t)(A^2  + t^p )^{5/2} }} {{B^5 }} - f(t)} \right]
\label{Case2c}
\end{eqnarray}

\item
Finally, in the last case, equations for $\lambda(t)$ and $y(t)$ were 
calculated in order to generate a wavefunction with oscillatory width.
The corresponding functions in this case read
\begin{eqnarray}\
y(t) &=& b + \sin ^2 (ct)
\label{Case3a}
\\
\lambda(t) &=& \frac{{f(t) - z(t)}}{{2\left[ {b + \sin ^2 (ct)} \right]^2 }}
\label{Case3b}
\\
z(t) &=&  - \frac{{c^2 }}{4}\sin ^2 (2ct) + c^2 \cos (2ct) \left[ {b + \sin ^2 (ct)} \right]
\label{Case3c}
\end{eqnarray}

\end{enumerate}

\begin{figure}[h]
\includegraphics[width=5cm,height=5cm,angle=0,clip]{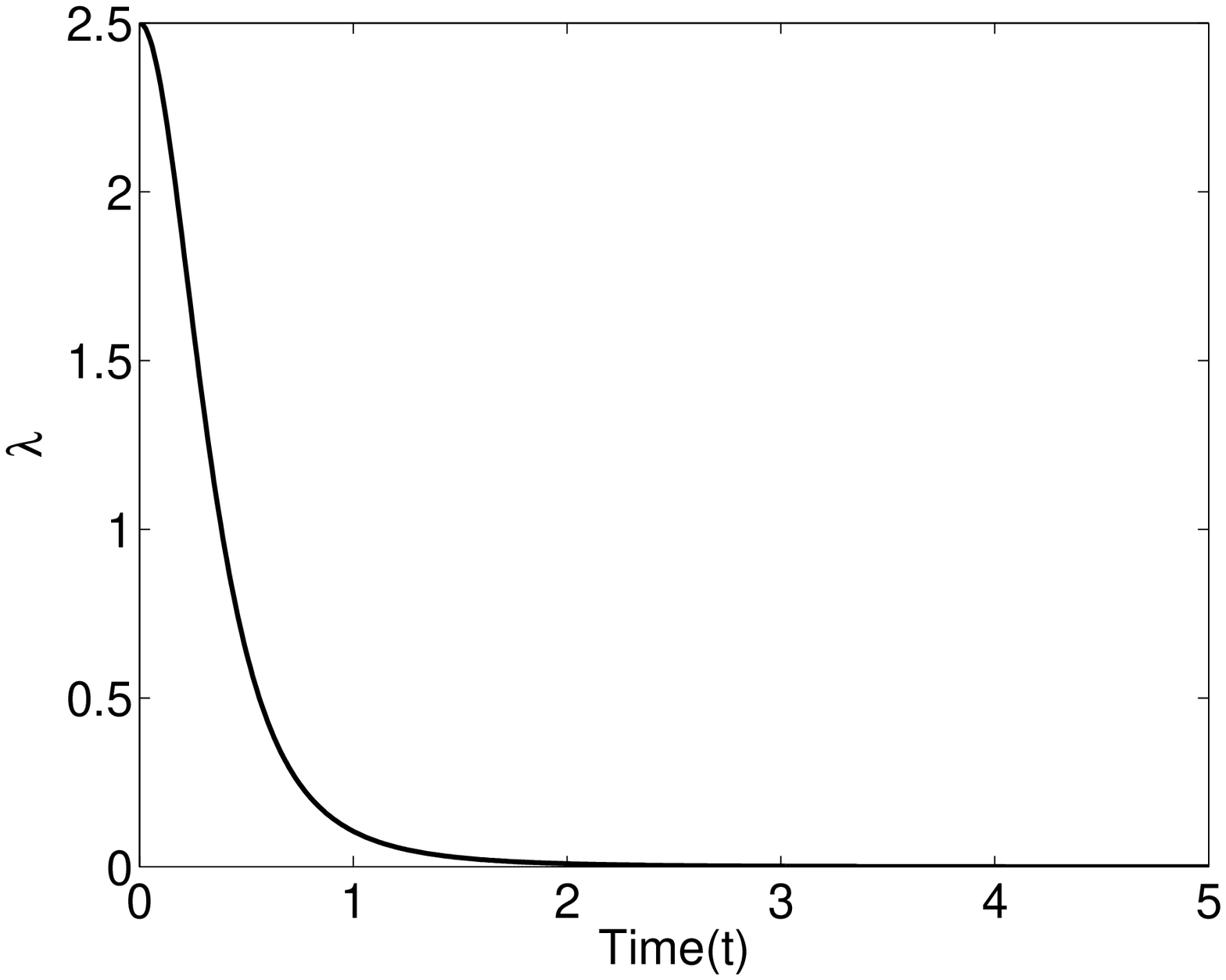} 
\includegraphics[width=5cm,height=5cm,angle=0,clip]{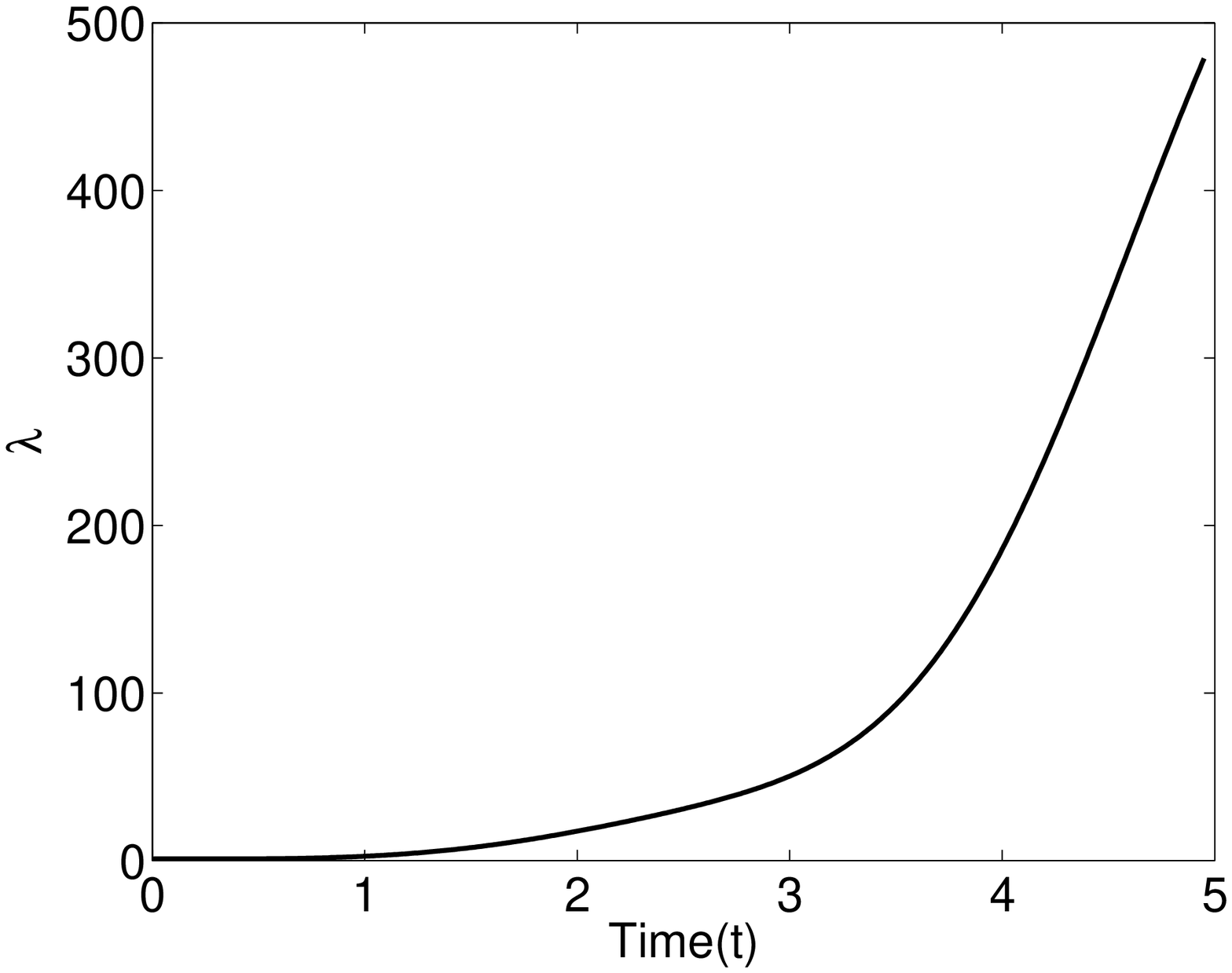} \includegraphics[width=5cm,height=5cm,angle=0,clip]{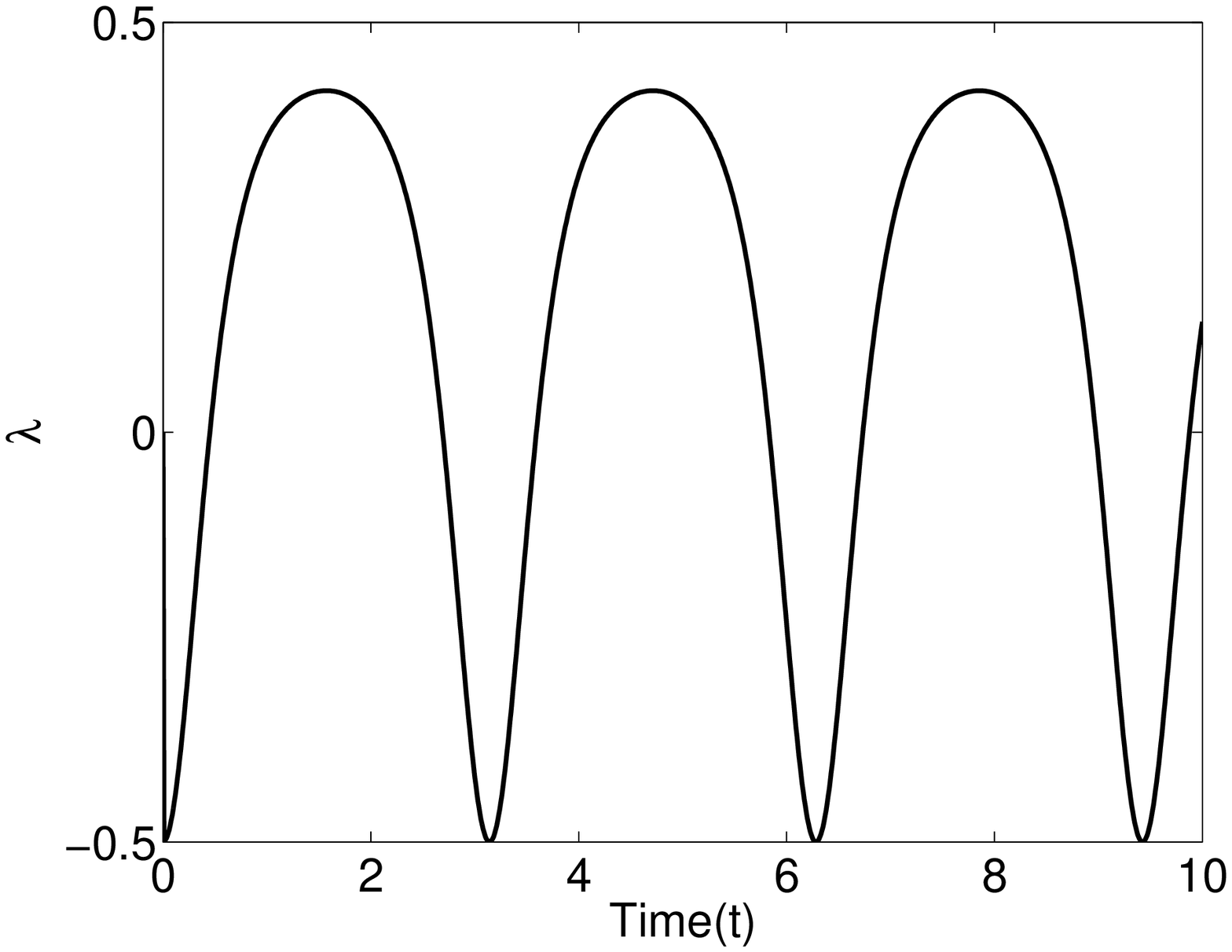}
\caption{Plots of $\lambda(t)$ for Cases 1-3, i.e., respectively 
expanding (left), contracting (middle) and oscillatory (right) condensates.}
\label{FigLambda}
\end{figure}

The resulting form of the normalized confining frequency $\lambda(t)$ for the three different cases is
shown in Fig. \ref{FigLambda}. We note in passing that such schemes
are definitely realizable within the
unprecedented control that exists over the 
magnetic confinement of the condensates. A very recent example illustrating this point
can be found in the recent work of \cite{engelss}; 
this experiment clearly realizes
a form of the third type of confinement (an oscillatory one),
by using a periodic 
modulation of the transverse confinement of
the condensate, in order to
induce longitudinal oscillations and the emergence of Faraday
patterns. This directly shows that temporal modulation
of the trapping frequencies is feasible. What we propose here consists of
the three principal types of trapping
frequency dependence, i.e., monotonic decrease, monotonic increase
and oscillatory dependence, that allow an explicit analytical
handle on the dynamics of the BEC, in excellent agreement with
our numerical findings (see below). Hence, we expect that such schemes would
be directly applicable in experimental settings similar to those
of the above experiment.

\section{Numerical Setup and Results}

In order to validate our analytical results based on the predictions
made above, the evolution of the condensates under the time-dependent
trappings imposed by Eqs. (\ref{Case1b}), (\ref{Case2b}) and (\ref{Case3b}) 
were tested by numerically solving the original PDE model of 
Eq. (\ref{eq1}). The numerical simulations were run 
using spectral methods \cite{trefethen}, with a spatial grid composed of 
Chebychev nodes ($x_n = cos(\pi n/N)$ where $N$ is the total number of nodes). 
Since the Chebychev nodes are contained within the interval $[-1,1]$, the 
spatial variable was normalized in order for the solution $u(t)$ to be 
supported on the interval $[-1,1]$ from $t=0$ to some evolution horizon 
$t=T>0$. The temporal integration was implemented using a 4th-order 
Runga-Kutta scheme.

It is worth noting here that in all three cases examined, the choice of $\nu(t)$ was independent of the 
equations for $\lambda(t)$ and $y(t)$, thus 
$\nu(t)=\sin^2(t)$ was chosen to be used for all cases, in consonance
with the Feshbach resonance management scheme discussed in the Introduction;
in that sense, however, notice that the key temporal variation in the schemes presented
herein is that of the magnetic trapping frequency. As an initial 
condition to Eqn. (\ref{eq1}), we used a generic Gaussian profile
of the form $u(t)=A \exp(-Br^2/2)$ with the values of $A$ and $B$ subject to the 
condition $y(0)=A^2/(2 B^2)$. 

\begin{figure}[h]
\begin{tabular}{c|c c }
\includegraphics[width=5cm,height=5cm,angle=0]{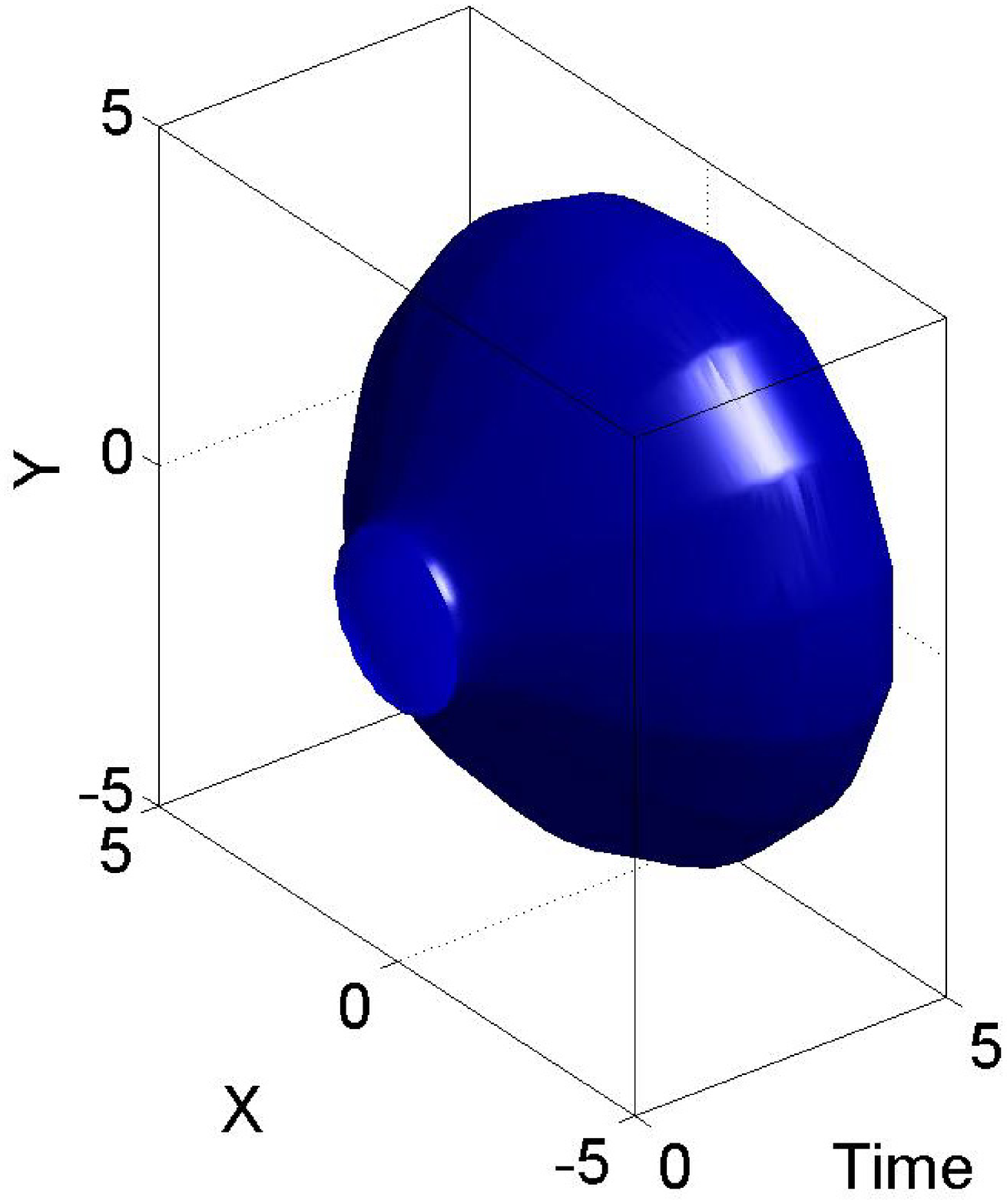} & \includegraphics[width=5cm,height=5cm,angle=0,clip]{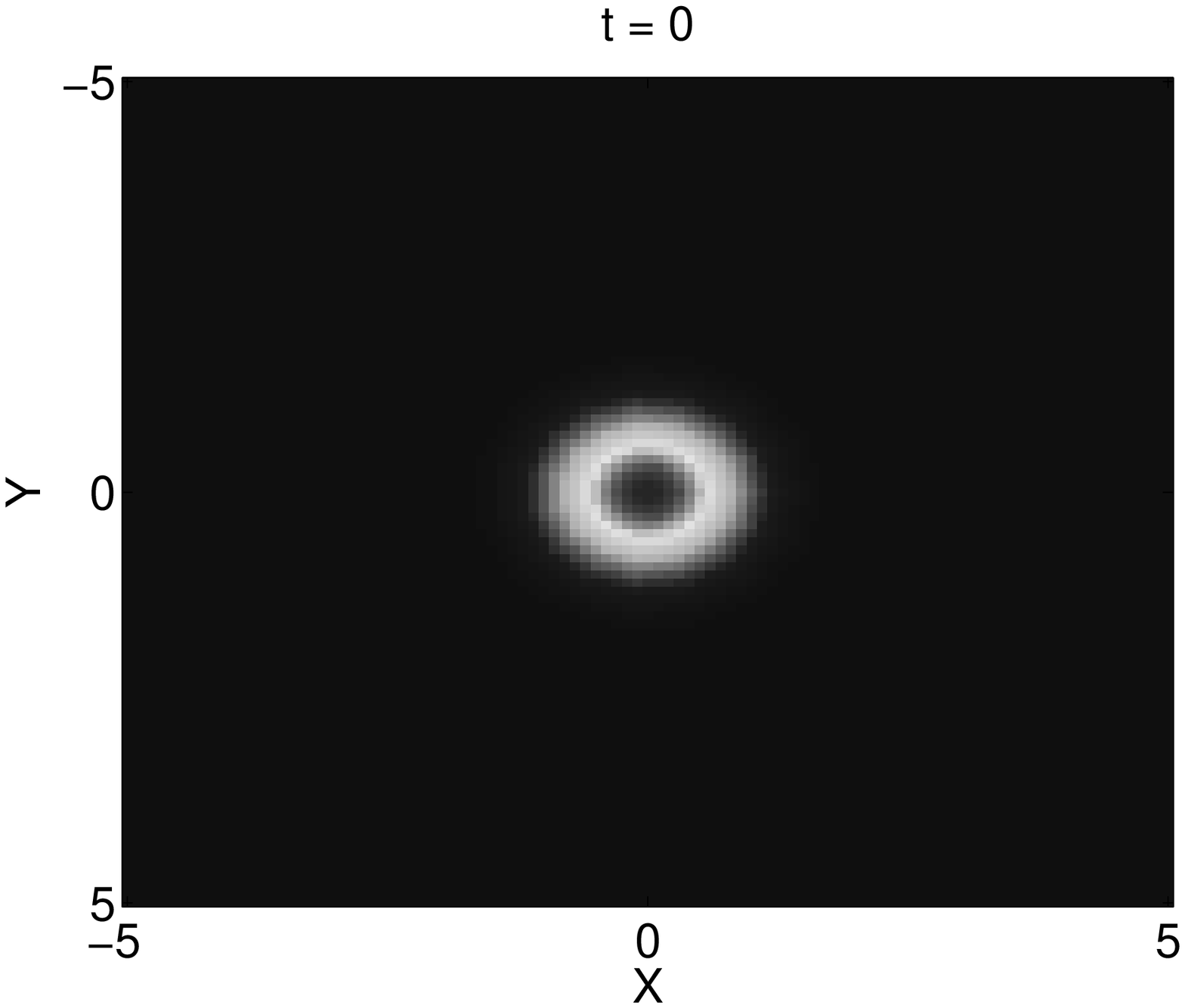} & \includegraphics[width=5cm,height=5cm,angle=0,clip]{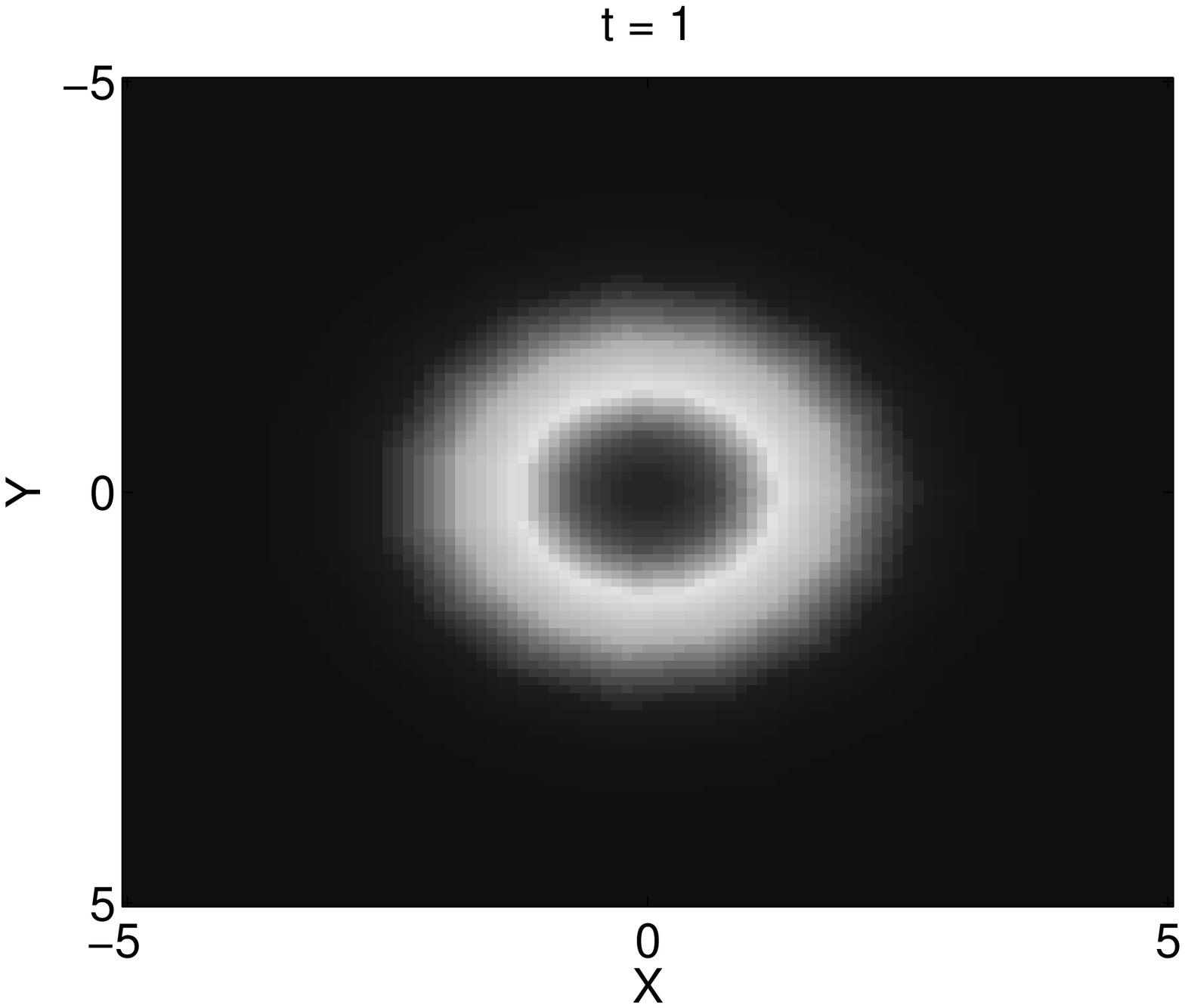}
\\
\includegraphics[width=5cm,height=5cm,angle=0,clip]{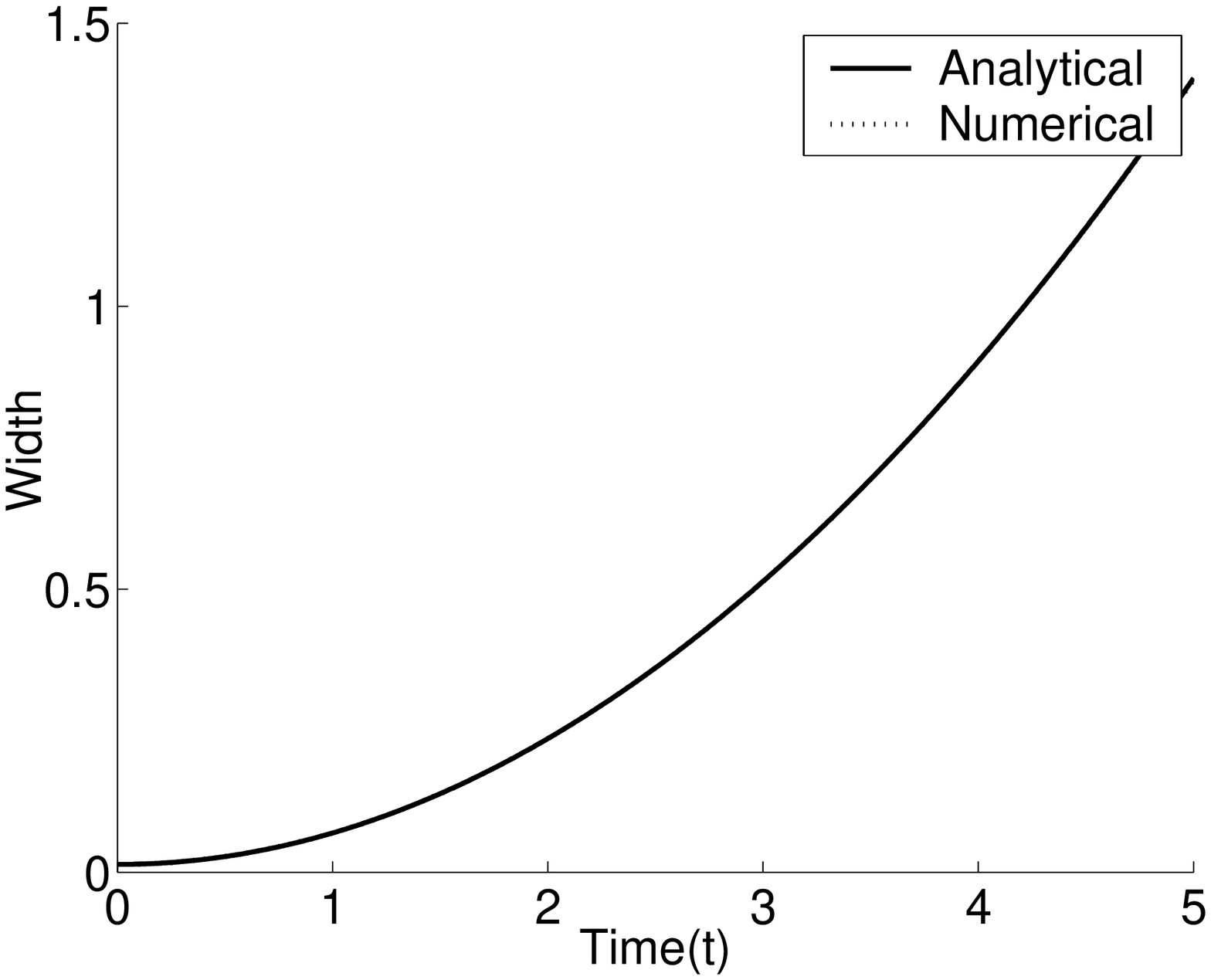} & \includegraphics[width=5cm,height=5cm,angle=0,clip]{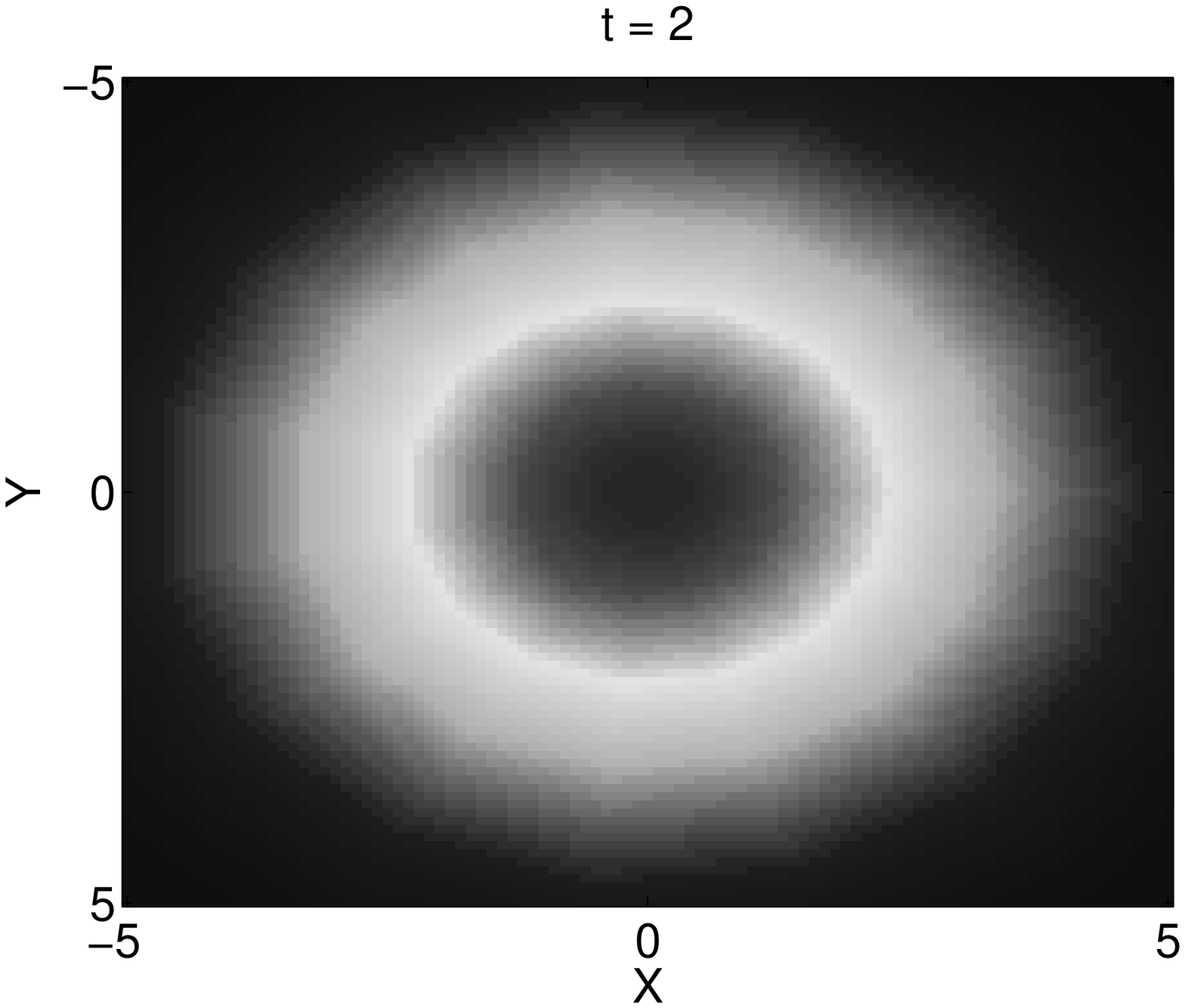} & \includegraphics[width=5cm,height=5cm,angle=0,clip]{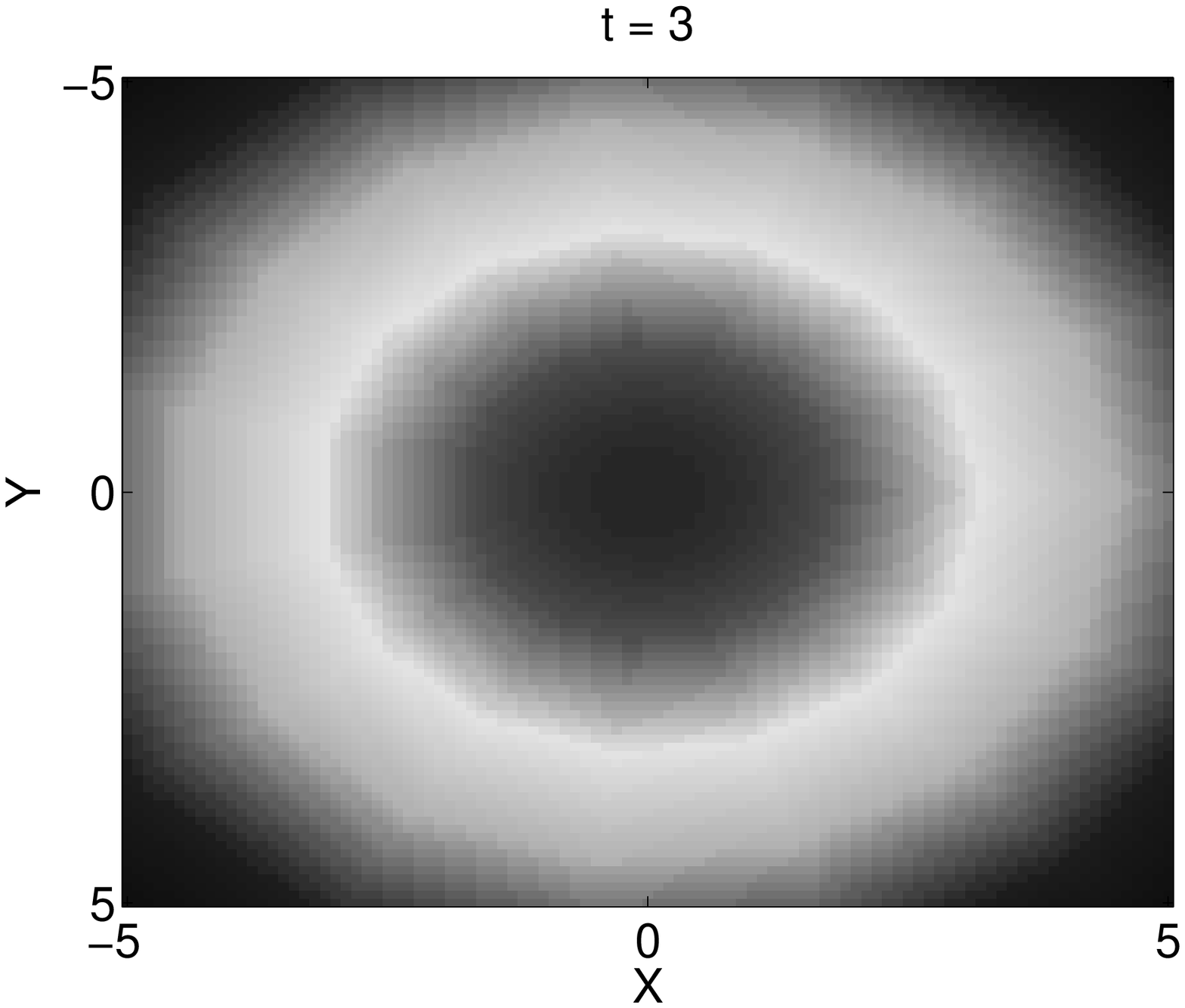}
\end{tabular}
\caption{Evolution plots of the expanding width wavefunction of Eqs. (\ref{Case1a})-(\ref{Case1b}). The upper left figure shows the spatio-temporal
evolution of a fixed contour of the solution over time. The lower left 
figure compares the analytically determined moment associated with 
the wavefunction width with its
numerically computed counterpart. The remaining figures show the 
contour of the solution at various times during the evolution, clearly
indicating the expanding nature of the BEC wavefunction.}
\label{FigCase1}
\end{figure}

\begin{figure}[h]
\begin{tabular}{c|c c }
\includegraphics[width=5cm,height=5cm,angle=0]{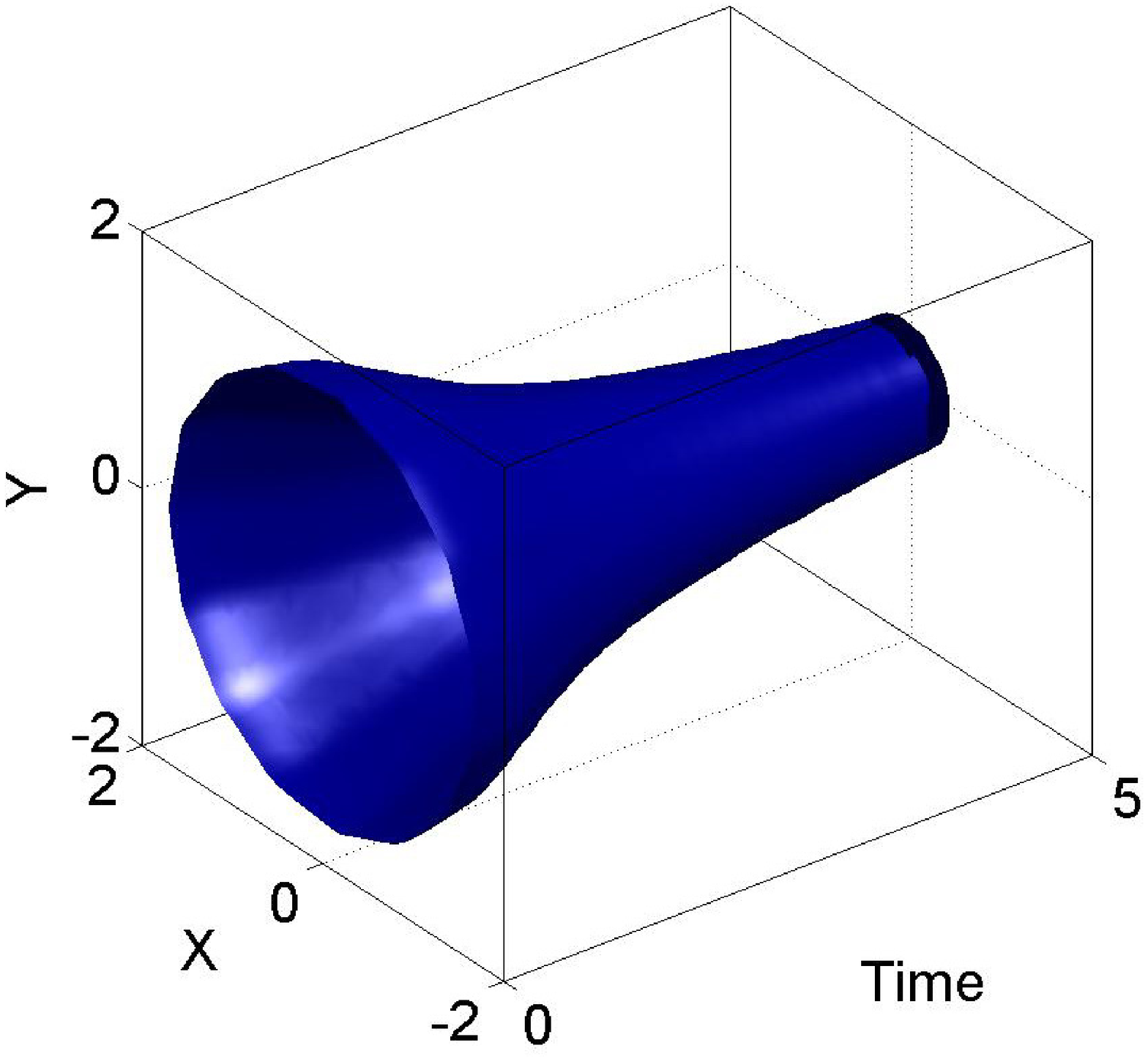} & \includegraphics[width=5cm,height=5cm,angle=0,clip]{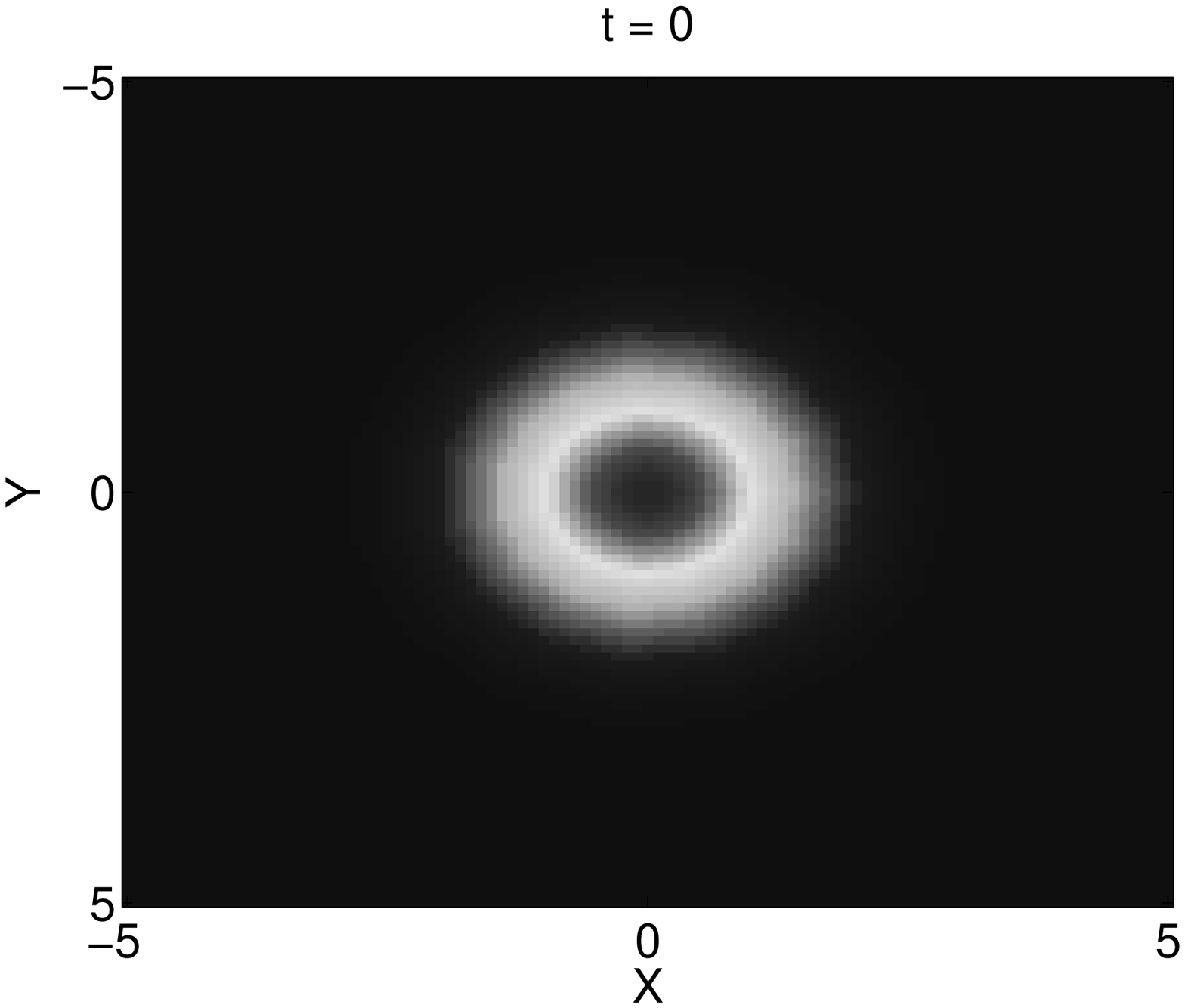} & \includegraphics[width=5cm,height=5cm,angle=0,clip]{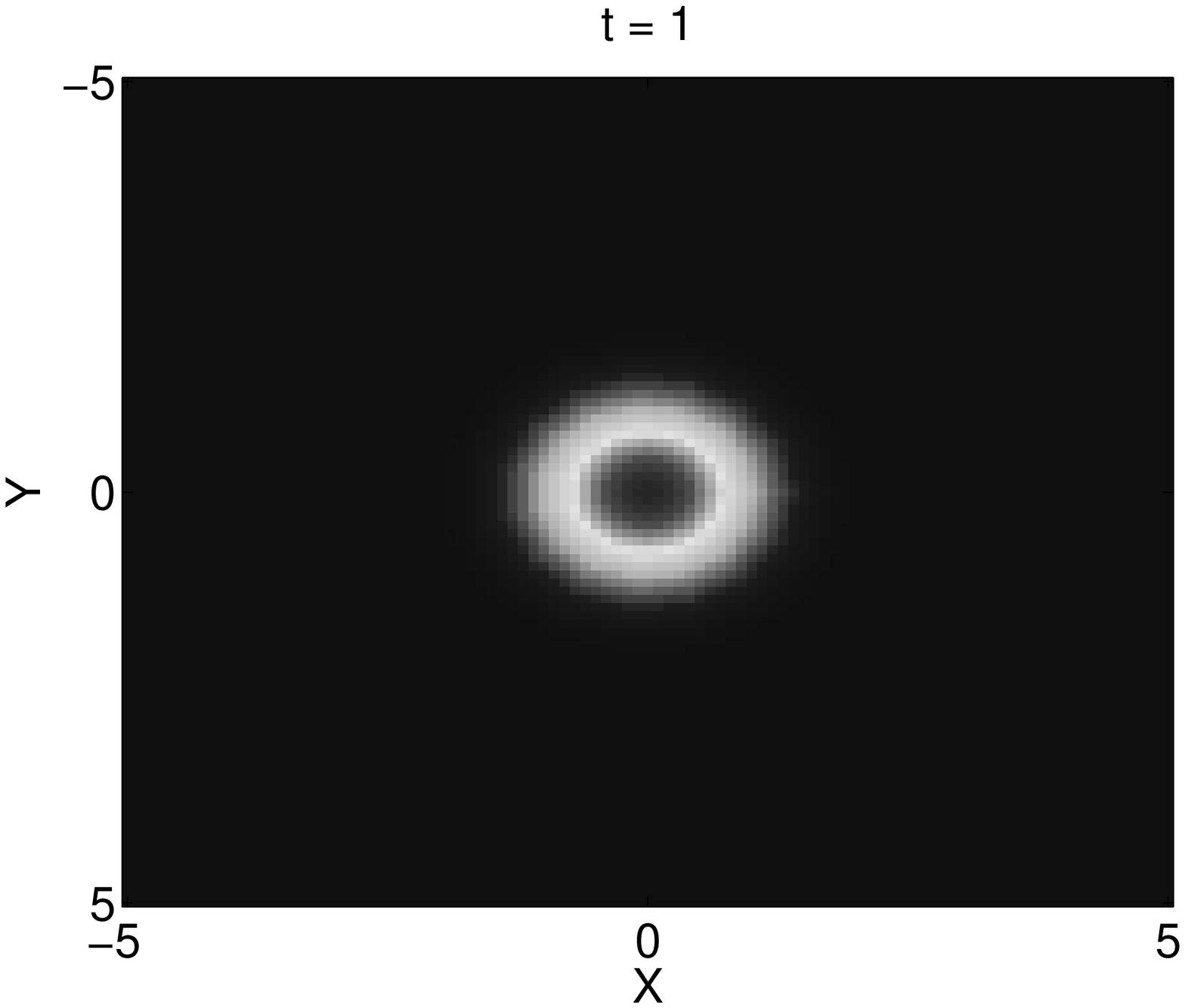}
\\
\includegraphics[width=5cm,height=5cm,angle=0,clip]{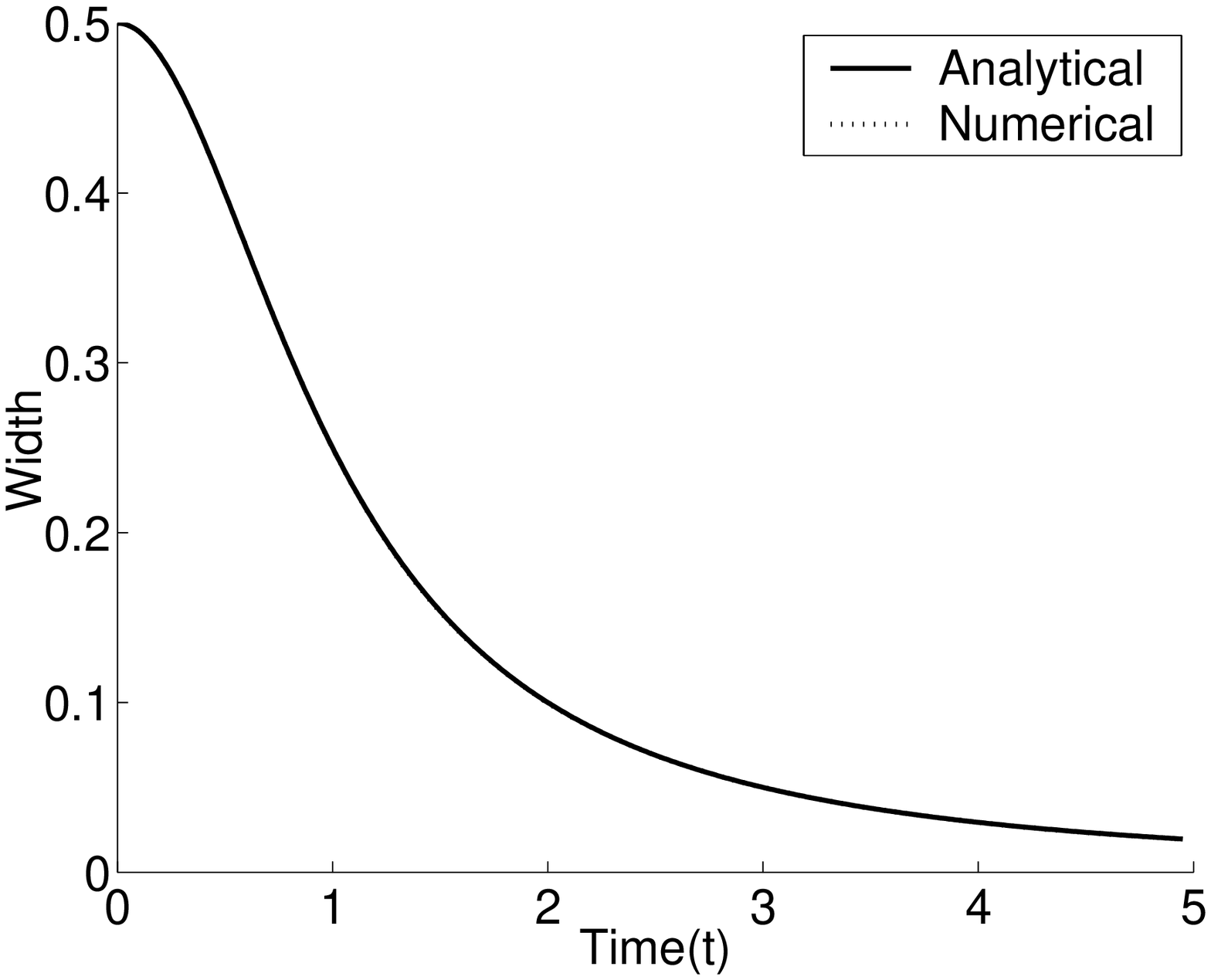} & \includegraphics[width=5cm,height=5cm,angle=0,clip]{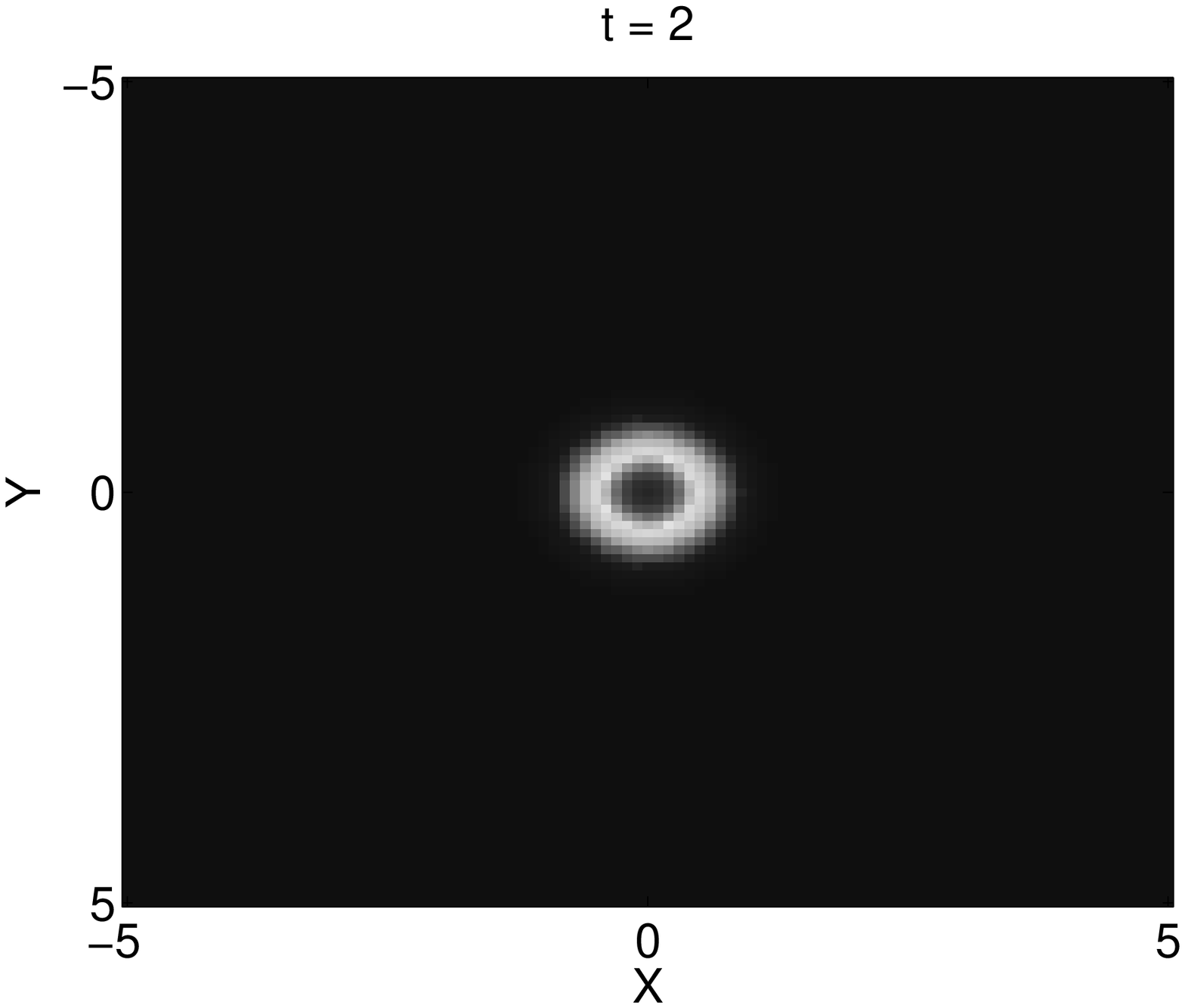} & \includegraphics[width=5cm,height=5cm,angle=0,clip]{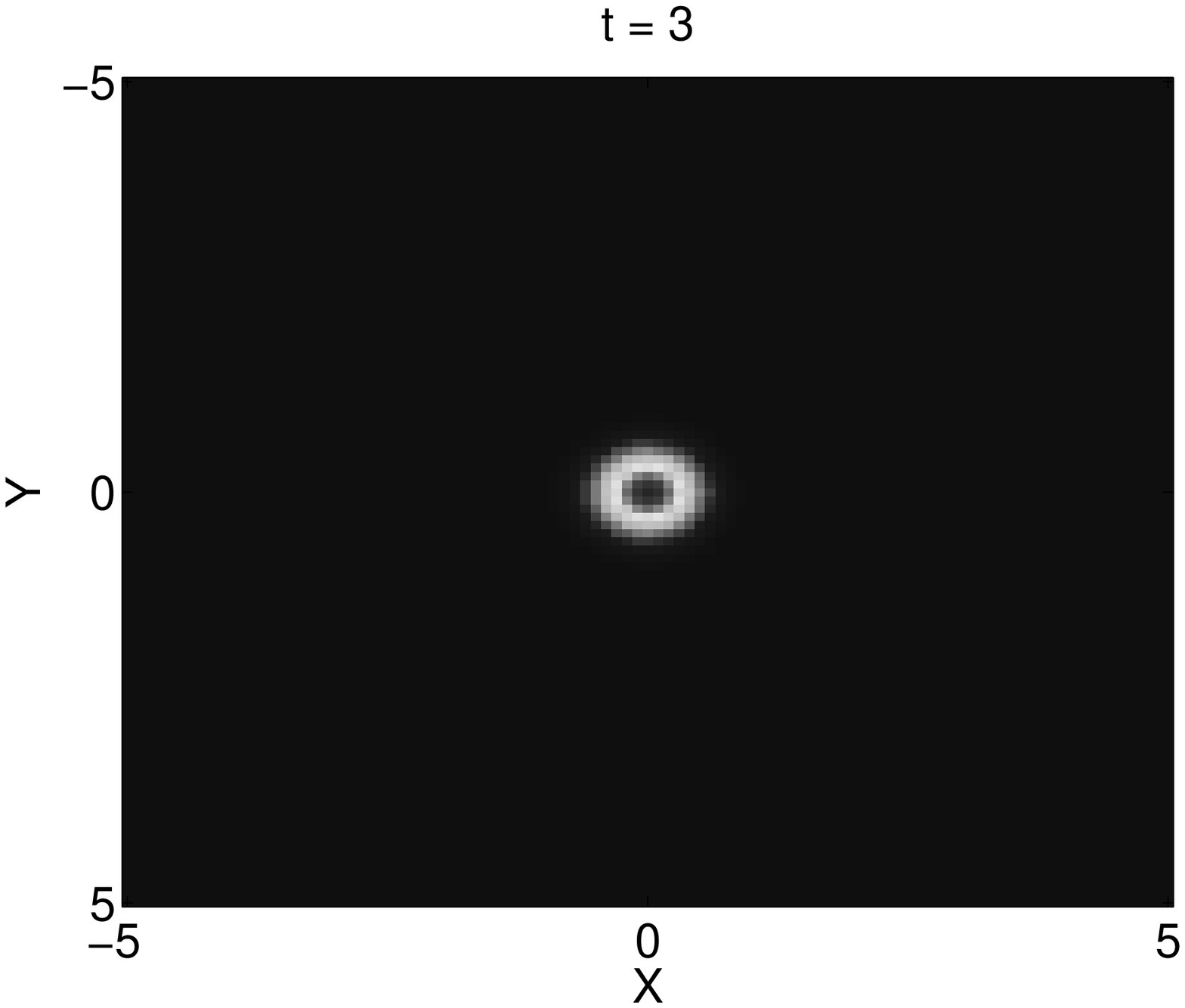}
\end{tabular}
\caption{Evolution plots of the focusing case of 
Eqs. (\ref{Case2a})-(\ref{Case2c}), with $p=2$ and $q=-9$. The upper left 
figure shows the spatio-temporal evolution of a fixed contour of the solution 
over time. The lower left figure compares the analytically determined 
wavefunction width with its numerically computed analog. 
The remaining figures show the contour of the solution at various times,
illustrating the focusing effect.}
\label{FigCase2}
\end{figure}

For the cases of the expanding and collapsing width wavefunctions, the 
numerical results are in perfect agreement with the analytical calculations 
as shown in Figs. (\ref{FigCase1}) and (\ref{FigCase2}). Notice in the 
lower left-hand plot for both figures how the lines for the analytical and 
numerical plots are completely indistinguishable from each other.

\begin{figure}[h]
\begin{tabular}{c|c c }
\includegraphics[width=5cm,height=5cm,angle=0]{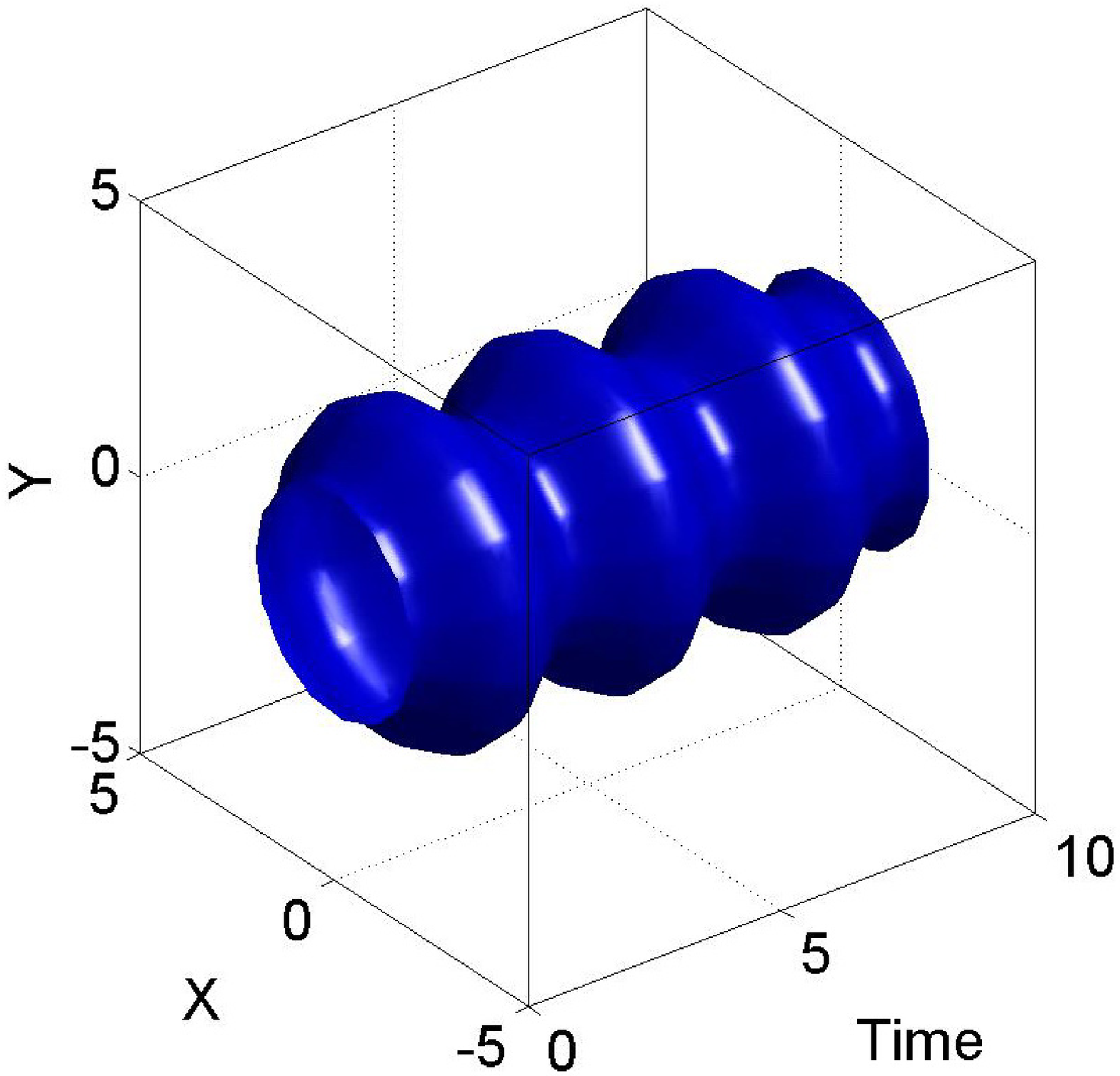} & \includegraphics[width=5cm,height=5cm,angle=0,clip]{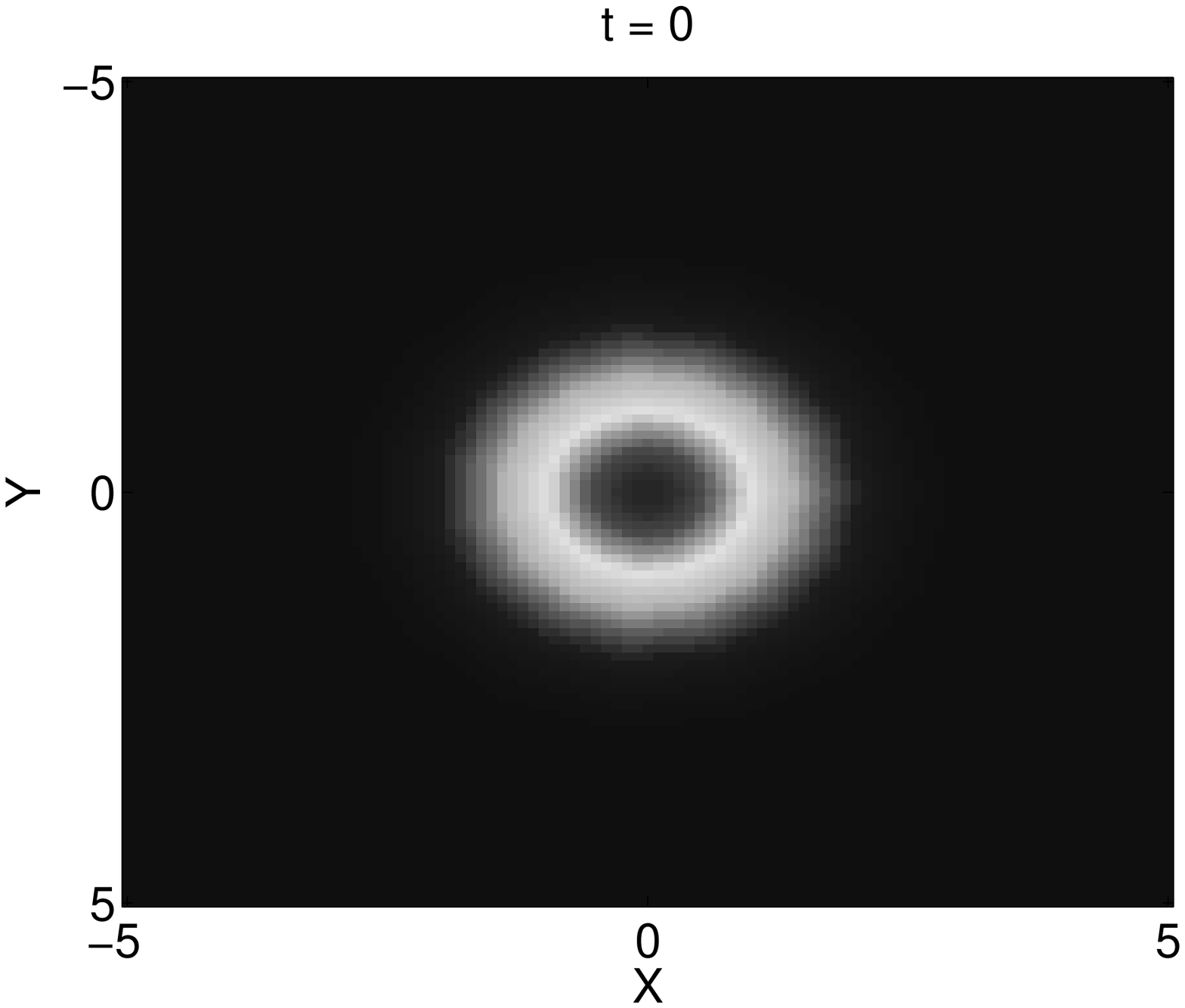} & \includegraphics[width=5cm,height=5cm,angle=0,clip]{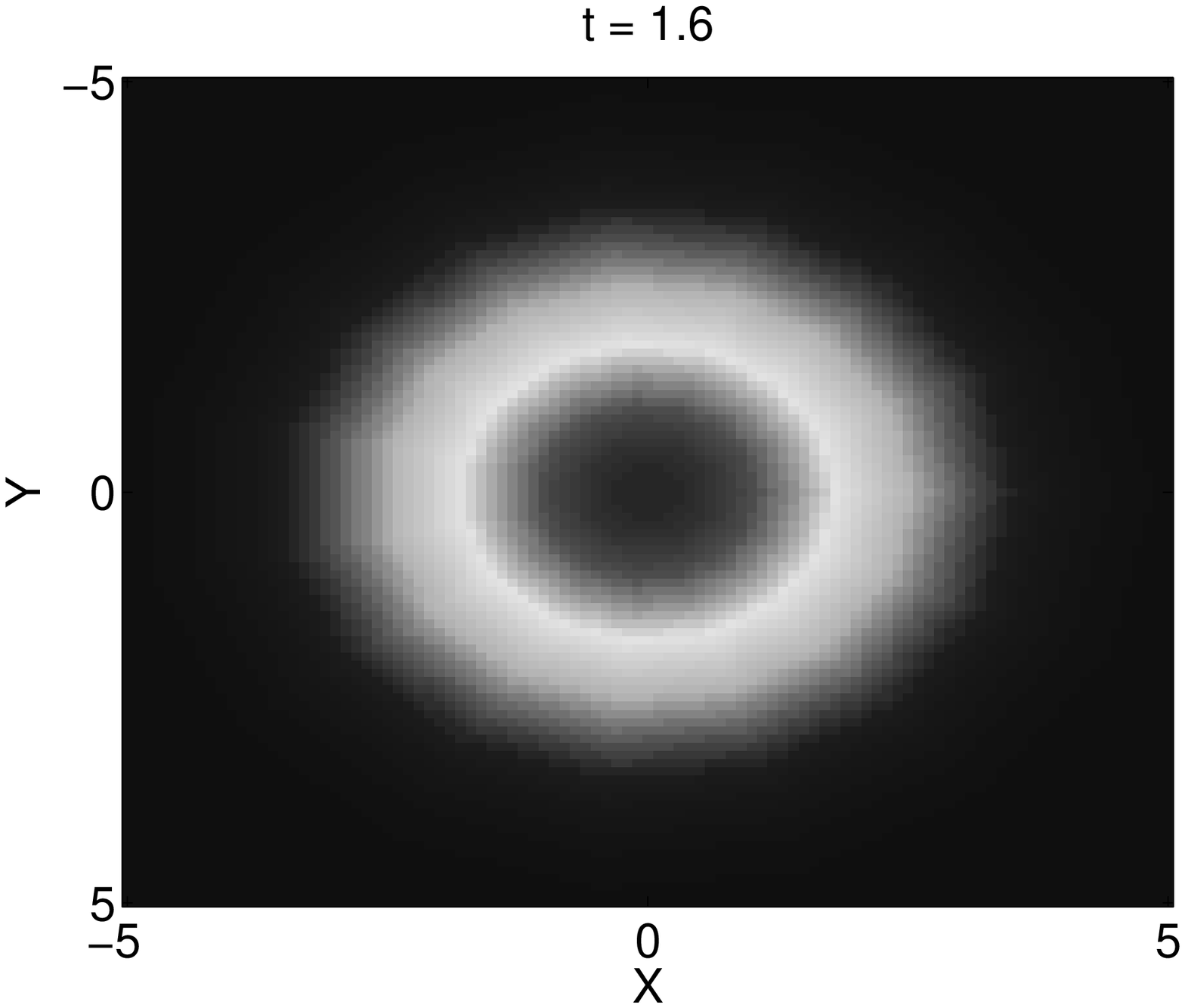}
\\
\includegraphics[width=5cm,height=5cm,angle=0,clip]{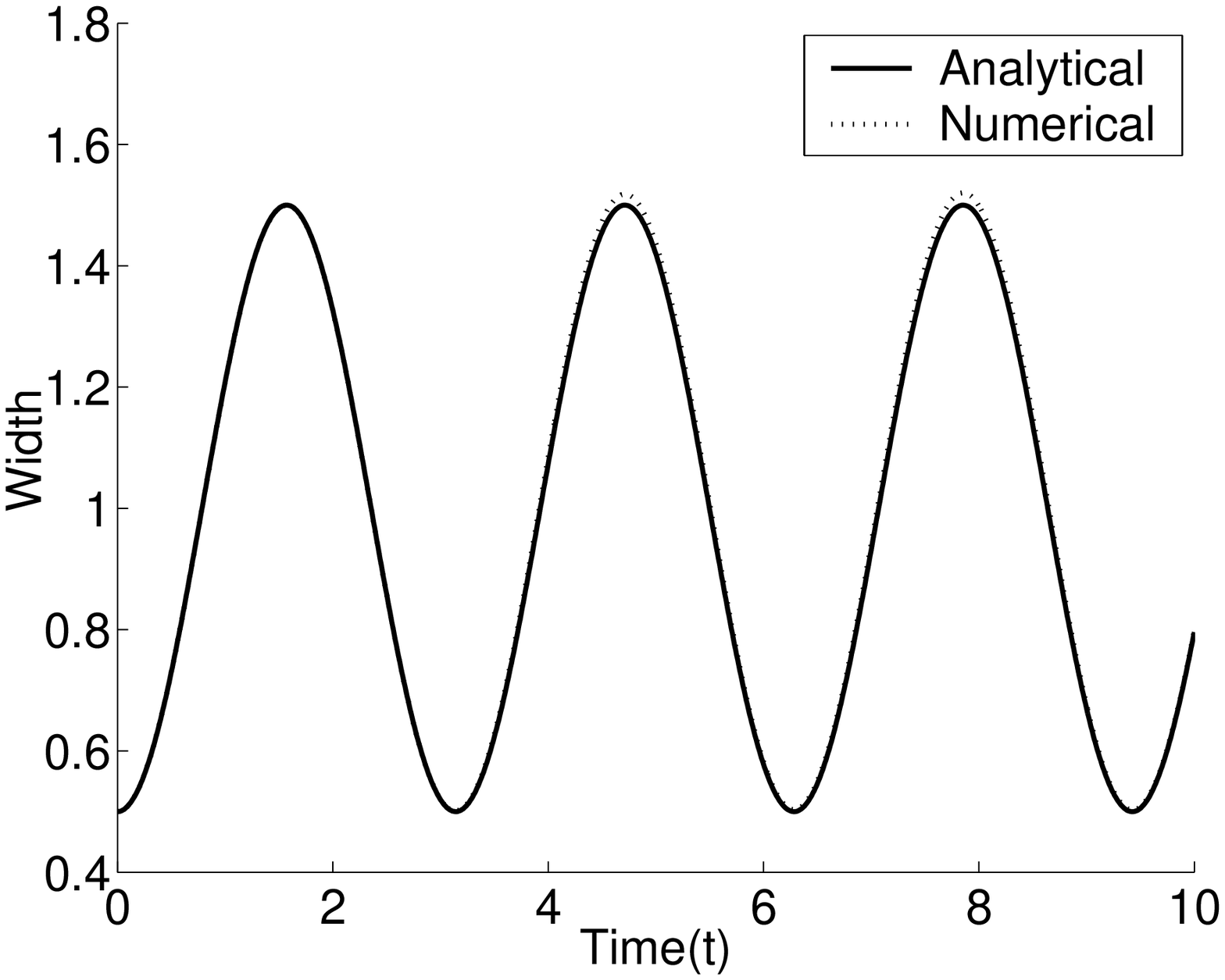} & \includegraphics[width=5cm,height=5cm,angle=0,clip]{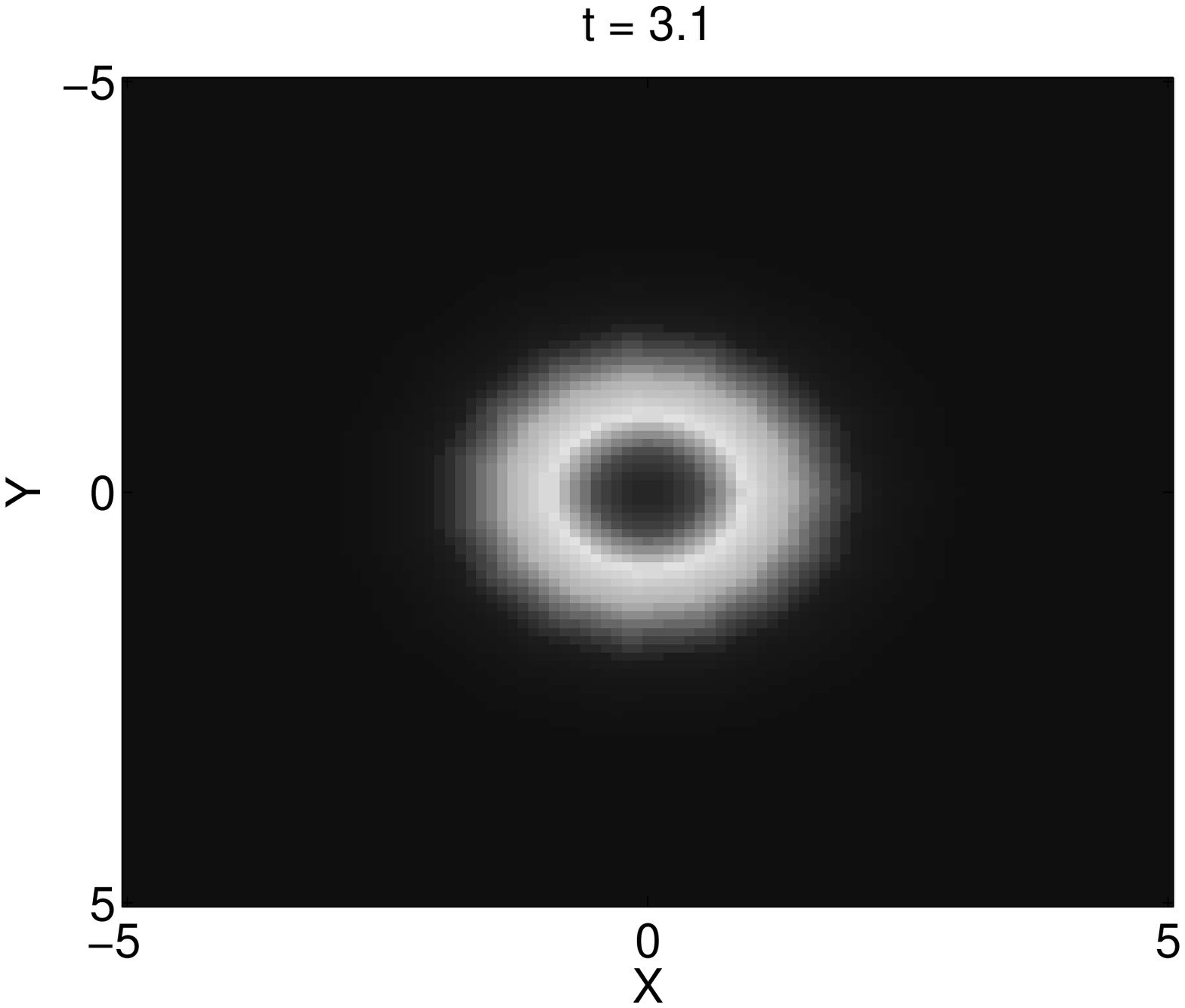} & \includegraphics[width=5cm,height=5cm,angle=0,clip]{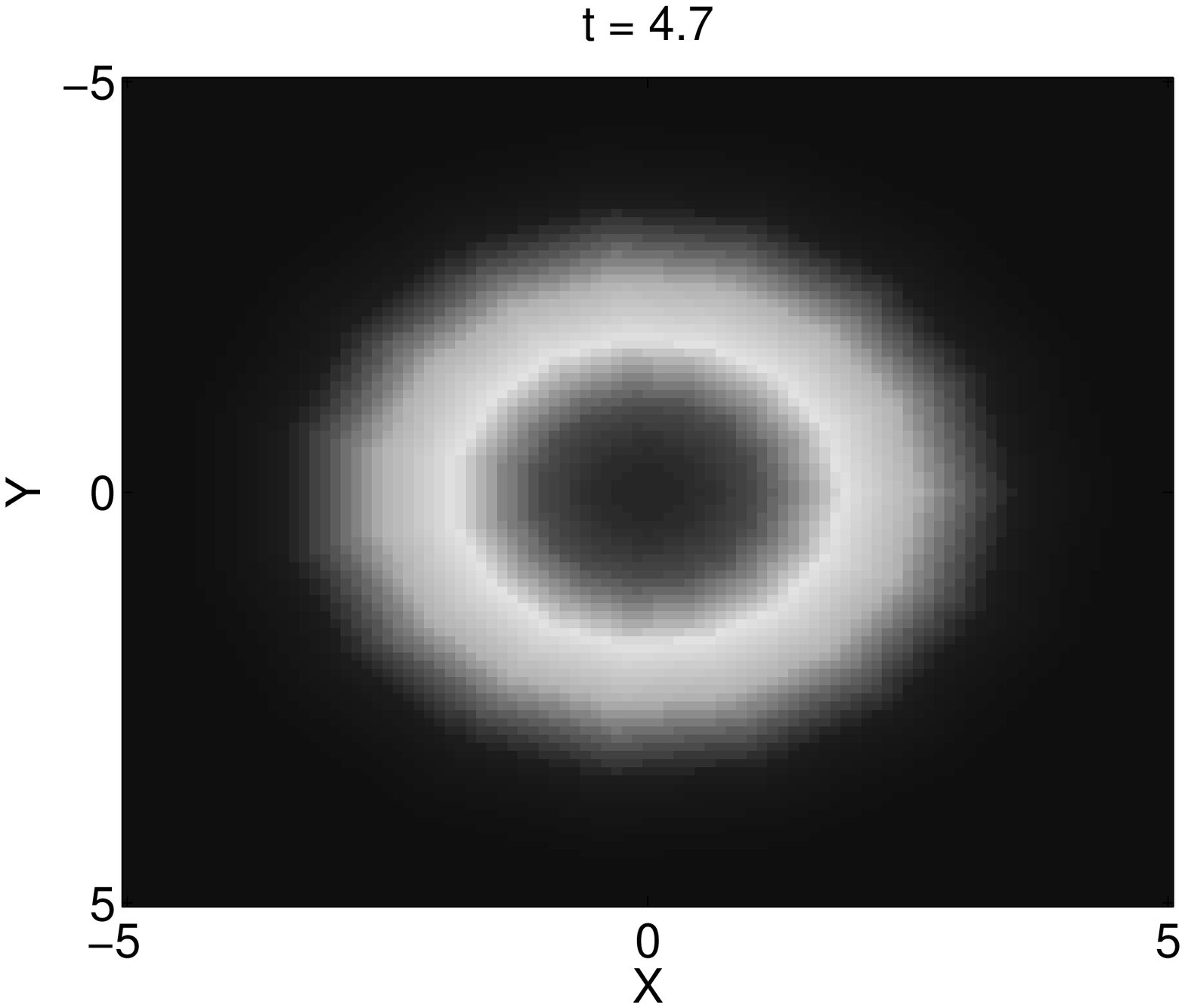}
\end{tabular}
\caption{Evolution plots for the oscillatory width wavefunction case 
of Eqs. (\ref{Case3a})-(\ref{Case3c}). 
The panels are similar to the previous two figures. 
The middle and right panels show the contour of the solution at
the first two maxima and minima of the wavefunction width.
}
\label{FigCase3}
\end{figure}

For the last case of the oscillatory width wavefunction, there was a very
small  discrepancy between the analytical calculations and numerical 
simulation, as seen in the lower left-hand plot of Fig. (\ref{FigCase3}). 
This discrepancy is only seen at the maximum values of the wavefunction width, 
starting with the second and subsequent maximums. 
While this small discrepancy may be triggered by the sole approximation
of our approach, namely Eq. (\ref{proportionality}), further numerical
experiments (not shown here) seem to indicate that it is more likely to
be the result of the numerical approximation to the corresponding moment.

In any case, the overall excellent numerical agreement between the
analytically obtained moment $I_{2,2}^{(2)}$ and its numerically
found counterpart illustrate the relevance of this approach and
its usefulness in systematically prescribing the wavefunction
behavior, in an analytically tractable way.


\section{Anisotropic Scalar Field Cosmologies}

An interesting analogy has been recently made between Bose-Einstein
condensates and isotropic scalar field cosmologies in the absence
of time dependence in the coefficient of the nonlinearity \cite{hawkins}. 
The so-called Friedmann-Robertson-Walker (FRW) metric,
coupled to a scalar field had been shown earlier \cite{hawk,floyd} to 
be described by an EP equation and that viewpoint was exploited in 
\cite{hawkins} to illustrate an analogy between cosmic dynamics and
BECs.

Here, we extend this analogy to EP equations with time-dependent
nonlinearities. In particular, we show that anisotropic cosmic dynamics
can be parallelized with Feshbach-managed Bose condensates.

If we consider the spatially homogeneous, yet anisotropic geometry
of Bianchi type I, we have a line element of:
\begin{eqnarray}
ds^{2}=-N(t)^{2}dt^{2}+ A(t)^2 dx^2 + B(t)^2 dy^2 + \Gamma(t)^2 dz^2.
\label{ceqn1}
\end{eqnarray}


We can define the tensor: $F_{\mu\nu}=G_{\mu\nu}-8\pi T_{\mu\nu}$
where $G_{\mu\nu}=R_{\mu \nu}-\frac{1}{2} g_{\mu \nu} R$ is the Einstein 
tensor and $T_{\mu \nu}=\phi_{;\mu}\phi_{;\nu}
-\frac{1}{2}g_{\mu\nu}(\phi^{;\alpha}\phi_{;\alpha}+m^{2}\phi^{2})$
is the energy momentum tensor. Then, the quadratic constraint
is the equation $F_0^0=0$, the kinematic equation is given
by $F_1^1=0$, while the Klein-Gordon equation for the
scalar field (that is coupled to gravity) 
is given by $\phi^{;\mu}_{;\mu}-m^{2}\phi=0 \propto
T^{\mu\nu}_{;\nu}=0$.
Notice additionally, that the two integrals of the motion,
namely $I_1=F^{1}_{1}-F^{2}_{2}=0$ and $I_2=F^{1}_{1}-F^{3}_{3}=0$
yield $B(t)=A(t)e^{\kappa
t/2}$ and $\Gamma(t)=A(t)e^{\lambda t/2}$.

Solving the Klein-Gordon equation for $\phi''(t)$ and substituting
the result into $\partial_t F_0^0=0$ (as well as solving $F_0^0=0$
for $\phi$ and using the resulting expression in $\partial_t F_0^0=0$),
one is led  to a dynamical equation for the remaining scale factor $A(t)$
in the form:
\begin{equation}
\frac{\kappa \,\lambda }{4} +
  \frac{\kappa \,A'(t)}{A(t)} +
  \frac{\lambda \,A'(t)}{A(t)} +
  \frac{4\,{A'(t)}^2}{{A(t)}^2} -
  \frac{{\phi '(t)}^2}{2} - \frac{A''(t)}{A(t)}=0
\label{gop1}
\end{equation}
Using now: $A(t)=Y(t)^{2/n}$ and a change of variable
$\tau=\int^t \Omega(t') dt'$, we obtain:
\begin{equation}
\ddot{Y}(\tau)+\dot{Y}(\tau)\frac{\Omega'(t)}{\Omega(t)^{2}}-\frac{\dot{Y}(\tau)^{2}}
{Y(\tau)}\frac{(6+n)}{n}-(\kappa+\lambda)\frac{\dot{Y}(\tau)}{\Omega(t)}
-\frac{n\kappa\lambda}{8}\frac{Y(\tau)}{\Omega(t)^{2}}+
\frac{n Y(\tau)\dot{\phi}(\tau)^{2}}{4}=0
\label{gop2}
\end{equation}
Hence, a choice of time reparametrization according to:
\begin{equation}
\frac{\Omega '(t)}{\Omega (t)} =
  \kappa  + \lambda  +\frac{(6+n)}{n}\frac{Y'(t)}{Y(t)}
\label{gop3}
\end{equation}
(which leads to $\Omega(t)=\theta e^{(\kappa+\lambda)t} Y(t)^{(6+n)/n}$,
where $\theta>0$ is a constant of integration),
results in the form:
\begin{equation}
\ddot{Y}(\tau)+ Q(\tau) Y(\tau)=\frac{\Gamma(t(\tau))}{Y(\tau)^{1+12/n}},
\quad Q=n\frac{\dot{\phi}(\tau)^{2}}{4}, \quad
\Gamma=\frac{n\kappa\lambda e^{-2(\kappa+\lambda)t}}{8\theta^{2}}.
\label{gop4}
\end{equation}
Eq. (\ref{gop4}) becomes
an
EP equation for the choice of $n=6$. Notice that in Eq. (\ref{gop4}),
a nontrivial complication is that
the time dependent coefficient $\Gamma$ depends on the reparametrization
of time through $\Omega$. Nevertheless, the original equation
(\ref{gop1}) has been solved in \cite{spanou}. This may, in turn,
provide valuable insights in the solution of such time-dependent
EP equations; this would be an interesting direction for future 
studies in such time-dependent EP settings.

We close this section by noting that one 
can therefore generalize the analogy of Bose condensates and scalar
field cosmologies in the Feshbach-managed viz. anisotropic case of the
EP equation with time-dependent nonlinearity. The quantities that
now bear the analogy is the scale factor of Bianchi type I with
the second moment of the condensate wavefunction, the time-dependent
magnetic trap strength with the time-dependent scalar field and
finally the time dependent nonlinearity coefficient (i.e., the
scattering length in the condensate dynamics) is (indirectly)
connected with the reparametrization of time in the cosmological
problem.

\section{Conclusions and Future Challenges}

In this short communication, we revisited the theme of
higher dimensional Bose-Einstein condensates under the presence
of magnetic trapping and Feshbach Resonance management. We 
used the moment method to develop an Ermakov-Pinney ODE
for the second moment of the distribution function,
which is associated with the width of the condensate.
This EP equation is analytically tractable, in a number
of cases. In particular, one can ``reverse engineer'' 
magnetic trappings that will induce the expansion, contraction
or oscillatory behavior of the condensate at will. For such
scenaria, the relevant moment of the wavefunction can be 
obtained analytically and is found to be in excellent
agreement with our numerical simulations of the full original
PDE model. The approach permits a detailed analytical handle
on the behavior of higher dimensional condensates which is
usually quite difficult to acquire with different methods
based on nonlinear PDEs. It may also, in turn, permit to
appropriately craft experiments, based on the reshaping ``operation''
that is desirable to perform on the condensate.

As an aside example, we have illustrated the relevance of
such time-dependent Ermakov-Pinney equations in a completely
different physical system, namely in anisotropic scalar
field cosmologies of Bianchi type I. 

This approach also suggests a number of interesting questions.
It would be, in particular, relevant to examine whether a general
solution of the time-dependent EP equation developed herein 
can be obtained on the basis of its linear Schr{\"o}dinger 
counterpart. This would bear direct consequences both in 
the atomic physics and in the cosmological problem,
allowing for a general analytical description of their respective
time-dependent properties (for the condensate wavefunction width or the
cosmological scale factor, respectively).
Furthermore, the approach was presented here
for two-dimensional settings for reasons that have to do
with the simplification/closure of the moment approach in that
case. However, it would be of particular interest to develop
similar approaches and potential closure schemes for
one-dimensional or three-dimensional settings. Such 
studies are currently in progress and will be reported
in future publications.




\end{document}